\documentclass[11pt]{article}

\usepackage[ansinew]{inputenc}
\usepackage{amsfonts}
\usepackage{amssymb,amsmath}
\usepackage{latexsym}
\usepackage{graphics}
\usepackage{epic}
\usepackage{epsfig}
\usepackage{psfrag}
\usepackage{subfigure}
\usepackage{bbm}

\numberwithin{equation}{section}

\newcommand     {\NN}{\mathbb{N}}
\newcommand     {\RR}{\mathbb{R}}
\newcommand     {\PP}{\mathbb{P}}

\newcommand     {\EE}{\mathbb{E}}

\newtheorem     {thm}{Theorem}[section]

\newtheorem     {lem}[thm]{Lemma}
\newtheorem     {prop}[thm]{Proposition}
\newtheorem     {cor}[thm]{Corollary}

\newtheorem     {rem}[thm]{Remark}

\newcommand     {\indic}        [1]
{{\mathbbm{1}}_{\{#1\}}}
\newcommand     {\indicbis}     [1]
{{\mathbbm{1}}_{#1}}

\begin{document}

\title{Splitting trees with neutral Poissonian mutations I: Small families.} 

\author{\textsc{Nicolas Champagnat$^{1}$, Amaury Lambert$^{2}$}}

\footnotetext[1]{TOSCA project-team, INRIA Sophia Antipolis -- M\'editerran\'ee, 2004 route des Lucioles, BP.\
  93, 06902 Sophia Antipolis Cedex, France, E-mail: \texttt{Nicolas.Champagnat@sophia.inria.fr}}

\footnotetext[2]{Laboratoire de Probabilités et Modèles Aléatoires,
UMR 7599 CNRS and UPMC Univ Paris 06,
Case courrier 188, 
4 Place Jussieu,
F-75252 Paris Cedex 05, France, Email: \texttt{amaury.lambert@upmc.fr}
}
\maketitle

\begin{abstract}

We consider a neutral dynamical model of biological diversity, where individuals live and reproduce independently. They have i.i.d.  lifetime durations (which are not necessarily exponentially distributed) and give birth (singly) at constant rate $b$.
Such a genealogical tree is usually called a splitting tree \cite{GK}, and the population counting process $(N_t;t\ge 0)$ is a homogeneous, binary Crump--Mode--Jagers process.

We assume that individuals independently experience mutations at constant rate $\theta$ during their lifetimes, under the infinite-alleles assumption: each mutation instantaneously confers a brand new type, called allele, to its carrier.
We are interested in the allele frequency spectrum at time $t$, i.e., the number $A(t)$ of distinct alleles represented  in the population at time $t$, and more specifically, the numbers $A(k,t)$ of alleles represented by $k$ individuals at time $t$, $k=1,2,\ldots,N_t$.

We mainly use two classes of tools: coalescent point processes, as defined in \cite{L10}, and branching processes counted by random characteristics, as defined in \cite{J,JNa}. 
We provide explicit formulae for the expectation of $A(k,t)$ conditional on population size in a coalescent point process, which apply to the special case of splitting trees. We separately derive the a.s. limits of $A(k,t)/N_t$ and of $A(t)/N_t$ thanks to random characteristics, in the same vein as in \cite{Taib}.

Last, we separately compute the expected homozygosity by applying a method introduced in \cite{L09}, characterizing the dynamics of the tree distribution as the origination time of the tree moves back in time, in the spirit of backward Kolmogorov equations.
\end{abstract}          
\bigskip



\section{Introduction}
\label{sec:intro}

We consider a general branching population, where individuals reproduce independently of each other, have i.i.d.\ lifetime durations, and give birth at constant rate during their lifetime. We also assume that each birth gives rise to a single newborn. The genealogical tree associated with this construction is known as a splitting tree \cite{Geiger, GK, L10}.
The process $(N_t;t\ge 0)$ counting the population size is a non-Markovian birth--death process belonging to the class of general branching processes, or Crump--Mode--Jagers (CMJ) processes. Since births arrive singly and at constant rate, these processes are sometimes called homogeneous, binary CMJ processes. 

Next, individuals are given a type, called allele or haplotype. They inherit their type at birth from their mother, and (their germ line) can change type throughout their lifetime, at the points of independent Poisson point processes with rate $\theta$, conditional on lifetimes (neutral mutations). The type conferred by a mutation is each time an entirely new type, an assumption known as the infinitely-many alleles model.

We are interested in the so-called allelic partition (partition into types) of the population alive at time
$t$. A convenient way of describing this partition without labelling types is to define the number
$A_\theta(k,t)$ of types carried by $k$ individuals at time $t$. The sequence $(A_\theta(k,t);k\ge 1)$ is
called the frequency spectrum of the allelic partition. We also denote by $A_\theta (t)$ the total number of
distinct types at time $t$. The most celebrated mathematical result in this setting is Ewens' sampling
formula, which yields the distribution of the frequency spectrum for the Kingman coalescent tree with neutral
Poissonian mutations \cite{Ewens}.

Credit is due to G.\ Yule \cite{Y} for the first study of a branching tree with mutations, but the interest for the infinitely-many alleles model applied to branching trees has started with the work of R.C.\ Griffiths and A.G.\ Pakes \cite{GP}, where the tree under focus is a Galton--Watson tree and each individual, with a fixed probability,  is independently declared mutant.  A fascinating monography dedicated  to general branching processes (also undergoing mutations at birth times) is due to  Z.\ Taïb \cite{Taib}. An extensive use is done there of a.s.\ limit theorems for branching processes counted by random characteristics, due to P.\ Jagers and O.\ Nerman \cite{J, JNa, JNb, N}.

More
recently, 
in a series of three companion papers, J.\ Bertoin~\cite{B1,B2,B3} has set up a very general framework for
Galton--Watson processes with mutations, where he has considered the allelic partition of the whole population
from origination to extinction, and studied various scaling limits for large initial population sizes and low
mutation probabilities.  Branching processes have also been used in the study of multistage carcinogenesis. In
this setting, the emphasis is put on the waiting time until a target mutation occurs, see \cite{DM, SS} and
the references therein.


In this paper, we study the part of the frequency spectrum corresponding to families with a fixed number of carriers, that we call small families. We use three techniques: coalescent point processes, branching processes counted by random characteristics, and Kolomogorov-type equations as a function of the origination time of the tree.
In a companion paper~\cite{CL}, we will discuss the part of the frequency spectrum corresponding to the largest or/and oldest families (the age of a family being that of their original mutation).

\section{Model and statement of main results}
\label{sec:models}

\subsection{Model}

In this work, we consider genealogical trees satisfying the branching property and called \emph{splitting trees} \cite{Geiger, GK}. Splitting trees are those random trees where individuals' lifetime durations are i.i.d.\ with an arbitrary distribution, but where birth events occur at Poisson times during each individual's lifetime. We call $b$ this constant birth rate and we denote by $V$ a r.v.\ distributed as the lifetime duration. Then set $\Lambda(dr):=b\PP(V\in dr)$ a finite measure on $(0,\infty]$ with total mass $b$ called the \emph{lifespan measure}. We will always assume that a splitting tree is started with one unique progenitor born at time 0.

The process $(N_t;t\ge0)$ counting the number of alive individuals at time $t$ is a homogeneous, binary \emph{Crump--Mode--Jagers process}, which is not Markovian unless $\Lambda$ has an exponential density or is the Dirac mass at $\{+\infty\}$.

\begin{figure}[ht]

\unitlength 2mm 
\linethickness{0.4pt}

\begin{picture}(66,33)(-5,10)
\put(4,39.875){\line(1,0){62}}
\put(10,40){\line(0,-1){9}}
\put(14,40){\line(0,-1){11.5}}
\put(18,40){\line(0,-1){4}}
\put(22,40){\line(0,-1){7}}
\put(26,40){\line(0,-1){16}}
\put(30,40){\line(0,-1){6}}
\put(34,39.875){\line(0,-1){8.5}}
\put(38,40){\line(0,-1){5.5}}
\put(42,40){\line(0,-1){11.5}}
\put(46,40){\line(0,-1){3.625}}
\put(50,40){\line(0,-1){22}}
\put(54,40){\line(0,-1){5}}
\put(58,40){\line(0,-1){7.5}}
\put(62,39.875){\line(0,-1){5}}
\put(6,40){\line(0,-1){25}}
\put(5.93,14.93){\line(0,-1){.8}}
\put(5.93,13.33){\line(0,-1){.8}}
\put(5.93,11.73){\line(0,-1){.8}}
\put(9.93,30.93){\line(-1,0){.8}}
\put(8.33,30.93){\line(-1,0){.8}}
\put(6.73,30.93){\line(-1,0){.8}}
\put(13.93,28.43){\line(-1,0){.8889}}
\put(12.152,28.43){\line(-1,0){.8889}}
\put(10.374,28.43){\line(-1,0){.8889}}
\put(8.596,28.43){\line(-1,0){.8889}}
\put(6.819,28.43){\line(-1,0){.8889}}
\put(17.805,35.93){\line(-1,0){.8}}
\put(16.205,35.93){\line(-1,0){.8}}
\put(14.605,35.93){\line(-1,0){.8}}
\put(21.93,32.93){\line(-1,0){.8889}}
\put(20.152,32.93){\line(-1,0){.8889}}
\put(18.374,32.93){\line(-1,0){.8889}}
\put(16.596,32.93){\line(-1,0){.8889}}
\put(14.819,32.93){\line(-1,0){.8889}}
\put(25.93,23.93){\line(-1,0){.9524}}
\put(24.025,23.93){\line(-1,0){.9524}}
\put(22.12,23.93){\line(-1,0){.9524}}
\put(20.215,23.93){\line(-1,0){.9524}}
\put(18.311,23.93){\line(-1,0){.9524}}
\put(16.406,23.93){\line(-1,0){.9524}}
\put(14.501,23.93){\line(-1,0){.9524}}
\put(12.596,23.93){\line(-1,0){.9524}}
\put(10.692,23.93){\line(-1,0){.9524}}
\put(8.787,23.93){\line(-1,0){.9524}}
\put(6.882,23.93){\line(-1,0){.9524}}
\put(29.93,33.93){\line(-1,0){.8}}
\put(28.33,33.93){\line(-1,0){.8}}
\put(26.73,33.93){\line(-1,0){.8}}
\put(33.93,31.43){\line(-1,0){.8889}}
\put(32.152,31.43){\line(-1,0){.8889}}
\put(30.374,31.43){\line(-1,0){.8889}}
\put(28.596,31.43){\line(-1,0){.8889}}
\put(26.819,31.43){\line(-1,0){.8889}}
\put(37.93,34.43){\line(-1,0){.8}}
\put(36.33,34.43){\line(-1,0){.8}}
\put(34.73,34.43){\line(-1,0){.8}}
\put(41.93,28.43){\line(-1,0){.9412}}
\put(40.047,28.43){\line(-1,0){.9412}}
\put(38.165,28.43){\line(-1,0){.9412}}
\put(36.283,28.43){\line(-1,0){.9412}}
\put(34.4,28.43){\line(-1,0){.9412}}
\put(32.518,28.43){\line(-1,0){.9412}}
\put(30.636,28.43){\line(-1,0){.9412}}
\put(28.753,28.43){\line(-1,0){.9412}}
\put(26.871,28.43){\line(-1,0){.9412}}
\put(45.93,36.43){\line(-1,0){.8}}
\put(44.33,36.43){\line(-1,0){.8}}
\put(42.73,36.43){\line(-1,0){.8}}
\put(49.93,17.93){\line(-1,0){.9778}}
\put(47.974,17.93){\line(-1,0){.9778}}
\put(46.019,17.93){\line(-1,0){.9778}}
\put(44.063,17.93){\line(-1,0){.9778}}
\put(42.107,17.93){\line(-1,0){.9778}}
\put(40.152,17.93){\line(-1,0){.9778}}
\put(38.196,17.93){\line(-1,0){.9778}}
\put(36.241,17.93){\line(-1,0){.9778}}
\put(34.285,17.93){\line(-1,0){.9778}}
\put(32.33,17.93){\line(-1,0){.9778}}
\put(30.374,17.93){\line(-1,0){.9778}}
\put(28.419,17.93){\line(-1,0){.9778}}
\put(26.463,17.93){\line(-1,0){.9778}}
\put(24.507,17.93){\line(-1,0){.9778}}
\put(22.552,17.93){\line(-1,0){.9778}}
\put(20.596,17.93){\line(-1,0){.9778}}
\put(18.641,17.93){\line(-1,0){.9778}}
\put(16.685,17.93){\line(-1,0){.9778}}
\put(14.73,17.93){\line(-1,0){.9778}}
\put(12.774,17.93){\line(-1,0){.9778}}
\put(10.819,17.93){\line(-1,0){.9778}}
\put(8.863,17.93){\line(-1,0){.9778}}
\put(6.907,17.93){\line(-1,0){.9778}}
\put(53.93,34.93){\line(-1,0){.8}}
\put(52.33,34.93){\line(-1,0){.8}}
\put(50.73,34.93){\line(-1,0){.8}}
\put(57.93,32.43){\line(-1,0){.8889}}
\put(56.152,32.43){\line(-1,0){.8889}}
\put(54.374,32.43){\line(-1,0){.8889}}
\put(52.596,32.43){\line(-1,0){.8889}}
\put(50.819,32.43){\line(-1,0){.8889}}
\put(61.93,34.93){\line(-1,0){.8}}
\put(60.33,34.93){\line(-1,0){.8}}
\put(58.73,34.93){\line(-1,0){.8}}
\put(6,41){\makebox(0,0)[cc]{$0$}}
\put(10,41){\makebox(0,0)[cc]{$1$}}
\put(14,41){\makebox(0,0)[cc]{$2$}}
\put(18,41){\makebox(0,0)[cc]{$3$}}
\put(22,41){\makebox(0,0)[cc]{$4$}}
\put(26,41){\makebox(0,0)[cc]{$5$}}
\put(30,41){\makebox(0,0)[cc]{$6$}}
\put(34,41){\makebox(0,0)[cc]{$7$}}
\put(38,41){\makebox(0,0)[cc]{$8$}}
\put(42,41){\makebox(0,0)[cc]{$9$}}
\put(46,41){\makebox(0,0)[cc]{$10$}}
\put(50,41){\makebox(0,0)[cc]{$12$}}
\put(54,41){\makebox(0,0)[cc]{$13$}}
\put(58,41){\makebox(0,0)[cc]{$14$}}
\put(62,41){\makebox(0,0)[cc]{$15$}}
\end{picture}

\caption{ A coalescent point process for $16$ individuals, hence $15$ branches.}
\label{fig : coalpointproc}
\end{figure}
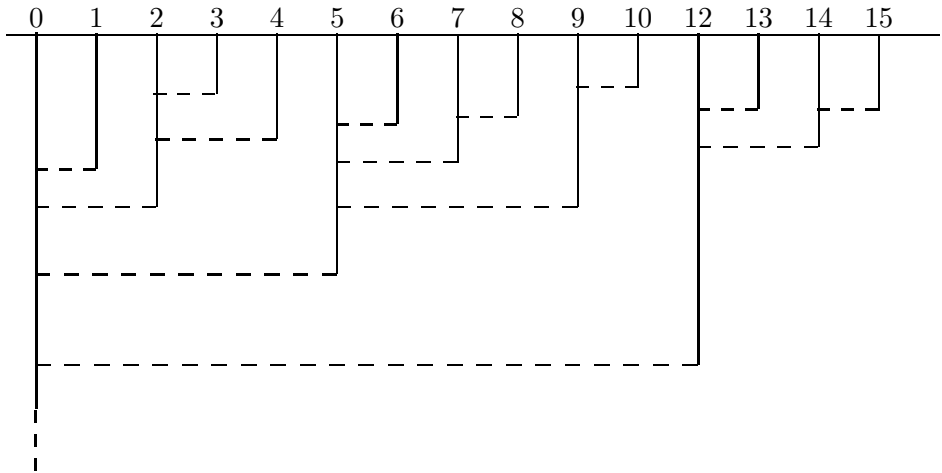

In \cite{L10}, it is shown that the genealogy of a splitting tree conditioned to be extant at a fixed time  $t$ is given by a \emph{coalescent point process}, that is, a sequence of i.i.d. random variables $H_i$, $i\ge 1$, killed at its first value greater than $t$. In particular, conditional on $N_t\not=0$, $N_t$ follows a geometric ditribution with parameter $\PP(H<t)$. More specifically, for any $0\le i \le N_t-1$, the \emph{coalescence time}  between the $i$-th alive  individual at time $t$ and the $j$-th individual alive at time $t$ (i.e., the time elapsed since the common lineage to both individuals split into two distinct lineages) is the maximum of $H_{i+1},\ldots, H_j$. The graphical representation on Figure \ref{fig : coalpointproc} is straightforward. The common law of these so-called \emph{branch lengths} is given by
\begin{equation}
  \label{eq:def-law H}
\PP(H>s)=\frac{1}{W(s)} , 
\end{equation}
where the nondecreasing function $W$ is such that $W(0)=1$ and is characterized by its Laplace transform. More
specifically, these branch lengths are the depths of the excursions of the jump contour process, say $Y^{(t)}$, of the splitting tree truncated below level $t$. They are i.i.d. because $Y^{(t)}$ is a Markov process. Indeed, it is shown in \cite{L10} that $Y^{(t)}$ has the law of a Lévy process, say $Y$, with no negative jumps, reflected below $t$ and killed upon hitting 0. The function $W$ is called the scale function of $Y$, and is defined from the Laplace exponent $\psi$ of $Y$:
\begin{equation}
  \label{eq:def-psi}
  \psi(x)=x-\int_{(0,+\infty]}\left(1-e^{-rx}\right) \Lambda (dr)\qquad x\in \RR_+.  
\end{equation}
Let $\alpha$ denote the largest root of $\psi$. In the supercritical case (i.e. $\int_{(0,\infty]} r\Lambda (dr)>1$), and in this case only, $\alpha$ is positive and called the \emph{Malthusian parameter}, because the population size grows exponentially at rate $\alpha$ on the survival event. Then the function $W$ is characterized by
$$
\int_0^\infty e^{-xr} W(r)\, dr = \frac{1}{\psi(x)}\qquad x>\alpha.
$$
Actually, it is possible to show by path decompositions of the process $Y$ that 
$$
W(x)=\exp\left( b\int_0^x dt\,\PP(J>t) \right),
$$
where $J$ is the maximum of the path of $Y$ killed upon hitting 0 and started from a random initial value, distributed as $V$. Note that since $Y$ is also the contour process of a splitting tree, $J$ has the law of the extinction time of the CMJ process $N$ started from one individual.

In the next section, we consider coalescent point processes without reference to a splitting tree. The law of such a process is merely characterized by a random number $N$ of i.i.d. r.v. $(H_i)$ independent of $N$, both with  arbitrary distributions. In this setting, \eqref{eq:def-law H} conversely serves as a \emph{definition of $W$}, which is now an arbitrary nondecreasing function, whereas it was previously seen to be differentiable in the special case of splitting trees.
The population size $N$ can be fixed (possibly infinite) or truly random, e.g. following a geometric distribution. It will be written $N_t$ when the law of $H$ is supported by $[0,t]$. In this latter case, any result obtained under the assumption that $N$ follows a geometric distribution can be applied to the case of splitting trees.

Throughout this work, we assume that individuals independently experience mutations at Poisson times during
their lifetime, that each new mutation event confers a brand new type (called haplotype, or allele) to the
individual, and that a newborn holds the same type as her mother at birth time. The mutation rate is denoted by $\theta$.

\subsection{Outline and statement of main results}

The main technique we use relies on the previously described representation of the genealogy of a splitting
tree by a sequence of i.i.d.\ {r.v.\ $(H_i)_{i\geq 1}$, called the coalescent point process (see also \cite{P}
for the critical, exponential case). The common distribution of $H_1, H_2, \ldots$ is related to the scale
function $W$.} We will also use the scale function $W_\theta$ associated with the lifetime of clonal families
(standard lifetime killed at its first mutation event). Section 3 is dedicated to some fine computations in
the general framework of coalescent point processes. For example, for a coalescent point process $(H_0,
H_1,\ldots, H_{X})$ of age $t$, where $X$ is an independent geometric r.v., Theorem \ref{thm : expected A
  geom} gives the expectation of $A_\theta(k,t)u^X$.  Various corollaries are stated, giving the expectation,
sometimes conditional on the population size, of specific quantities of biological interest at the fixed time
$t$. Those statements extend results of \cite{L09} given under a doubly asymptotic regime ($t,n\to\infty$).
For example, Corollary \ref{cor : IAG} gives the expectation of the number of distinct alleles and of
homozygosity (probability of drawing two individuals carrying the same allele) and Corollary \ref{cor : IZ}
gives the expectation of the number $Z_0(y;n)$ among the $n$ first individuals who carry the ancestral type of lineage 0 $y$ units of time in the past
$$
\EE \, Z_0(y;n)= e^{-\theta y}  \sum_{k=0}^n\PP(H\leq y)^k,
$$
see Remark \ref{rem : zzero} for a simple interpretation of this formula.

In Section 4, some of the previous results are specified to the case of splitting trees. In particular, Proposition \ref{prop : exp hf} yields the expectation of {$A_\theta(k,t)u^{N_t}$, as well as of $Z_0(t) u^{N_t}$,} where  $Z_0(t)$ denotes the number of alive individuals at time $t$ carrying the ancestral allele. The result for $A_\theta(k,t)$ can even be detailed to the case of haplotypes of a given age. As previously, various corollaries are provided for some quantities such as the homozygosity. Ruling out the information on the population size (i.e., taking $u=1$) and on the age of the mutation, Corollary \ref{cor : expected h freq} reads
$$
\EE^\star A_\theta (k,t) =W(t)\int_0^t dx\,\theta\,e^{-\theta x}\, \frac{1}{W_\theta(x)^2}\left(1-\frac{1}{W_\theta(x)}\right)^{k-1},
$$
and
$$
\PP^\star\left(Z_0(t)=k\right)= W(t)\,\frac{e^{-\theta t}}{W_\theta(t)^2}\left(1-\frac{1}{W_\theta(t)}\right)^{k-1},
$$
where $\PP^\star$ is the conditional probability on survival up until time $t$.
Note also that Subsection \ref{subsec : explanatory} provides the reader with a more explanatory proof of the previous formulae.

The theory of random characteristics~\cite{J,JNa,JNb,N, Taib}, which is the second main technique we use, is displayed in Section \ref{sec:a.s.-conv}. There, the random characteristic of individual $i$, say, can be for example the number $\chi_i^{k}(t)$ of mutations that $i$ has experienced during her lifetime and which are carried by  $k$ alive individuals, $t$ units of time after her birth ($\chi_i(t)=0$ if $t<0$). Then the total number of haplotypes carried by $k$ individuals at time $t$ (except possibly the ancestral type) is the sum over all individuals $i$ (dead or alive) of $\chi_i(t-\sigma_i)$, where $\sigma_i$ is the birth time of individual $i$. Now according to limit theorems by P.\ Jagers and O.\ {Nerman~\cite{J, JNa,JNb, N},} these sums converge a.s.\ on the survival event in the supercritical case.
Exploiting those limit theorems, we are able to independently derive the following a.s.\ convergences in the supercritical case (see Proposition  \ref{prop:a.s.-conv}). On the survival event,
$$
\lim_{t\to\infty} \frac{A_\theta(k,t)}{A_\theta(t)}=\frac{U_k}{U} \qquad a.s.
$$
and
$$
\lim_{t\to\infty} \frac{A_\theta(t)}{N_t}= U \qquad a.s.,
$$
where 
$$
U_k:=\int_0^\infty dx\,\theta\,e^{-\theta x}\, \frac{1}{W_\theta(x)^2}\left(1-\frac{1}{W_\theta(x)}\right)^{k-1},
$$
and
$$
U:=\sum_{k\ge 1}U_k=\int_0^\infty dx\,\theta\,e^{-\theta x}\, \frac{1}{W_\theta(x)} .
$$
In the final section (Section 6), we consider $G_\theta(t):=Z_0(t)(Z_0(t)-1)/2+\sum_{k\ge
    1}k(k-1)A_\theta(k,t)/2$, that we term absolute homozygosity, in reference to standard homozygosity,
which is defined as {$\bar{G}_\theta(t)=2G_\theta(t)/N_t(N_t-1)$.} Homozygosity is a well-known measure of diversity, that
can be seen as the probability that two randomly sampled {\emph{distinct}} individuals (or sequences) share
the same allele. In the spirit of backward Kolmogorov equations, we derive the dynamics of the expectation of
$G_\theta (t)u^{N_t}$ as the origination time of the tree moves back in time. Then the expected {standard and}
absolute homozygosity can be computed. In passing, we recover formulae obtained in
Section~\ref{sec:splitting-trees} by totally different methods. Specifically, we get $\EE^\star G_\theta(t)= W(t)
(W_{2\theta}(t) -1)$.

\section{Expected haplotype frequencies for coalescent point processes}
\label{sec:coalescent}

In this section, unless otherwise specified, we assume that the lineage of individual 0, sometimes called
lineage 0, is infinite, and that all other branch lengths are i.i.d., distributed as some r.v.\ $H$. To each
$H_i$ corresponds an individual, that we call individual $i$. We also
assume that mutations occur according to a Poisson point process on edge lengths with parameter $\theta$.

\subsection{The next branch with no extra mutation}

\label{subsec : next branch with no}
We let ${\cal E}^\theta$ denote the set of individuals who \emph{carry no more mutations} (but possibly less) than individual $0$
 (some of and at most exactly the mutations carried by $0$, but no other
mutation). We call such individuals \emph{$(0,\cdot)$-type individuals} (same type as some point on lineage 0
at some time in the past).

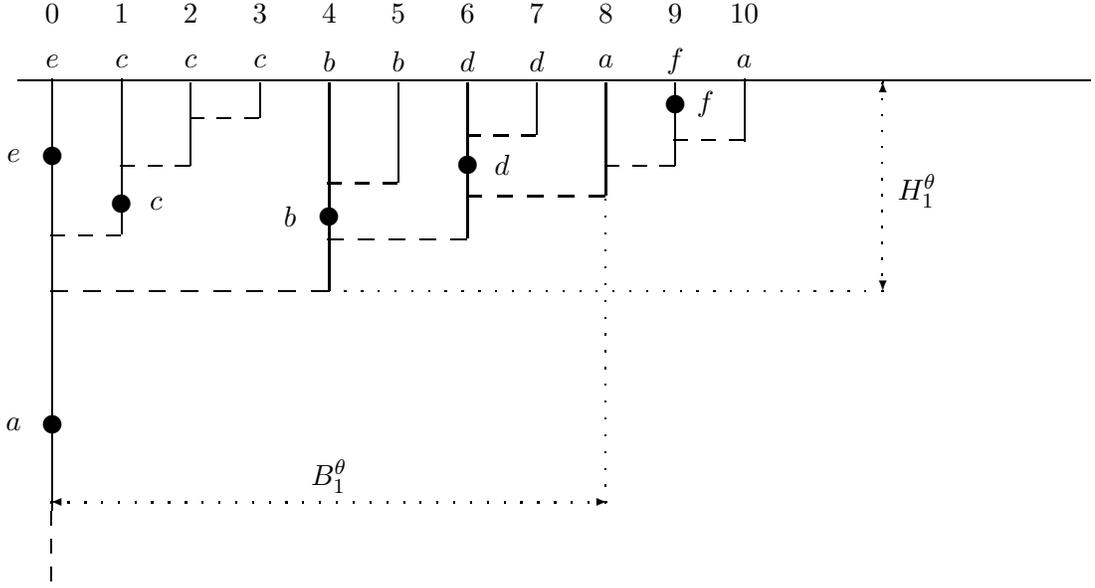
\begin{figure}[ht]
\unitlength 2.3mm 
\linethickness{0.4pt}

\ifx\plotpoint\undefined\newsavebox{\plotpoint}\fi 
\begin{picture}(71,33.75)(-5,10)
\put(4,39.875){\line(1,0){62}}
\put(10,40){\line(0,-1){9}}
\put(46,40){\line(0,-1){3.625}}
\put(6,40){\line(0,-1){25}}
\put(5.93,14.93){\line(0,-1){.8}}
\put(5.93,13.33){\line(0,-1){.8}}
\put(5.93,11.73){\line(0,-1){.8}}
\put(9.93,30.93){\line(-1,0){.8}}
\put(8.33,30.93){\line(-1,0){.8}}
\put(6.73,30.93){\line(-1,0){.8}}
\put(45.93,36.43){\line(-1,0){.8}}
\put(44.33,36.43){\line(-1,0){.8}}
\put(42.73,36.43){\line(-1,0){.8}}
\put(6,41){\makebox(0,0)[cc]{$e$}}
\put(6,43.75){\makebox(0,0)[cc]{$0$}}
\put(10,41){\makebox(0,0)[cc]{$c$}}
\put(10,43.75){\makebox(0,0)[cc]{$1$}}
\put(14,41){\makebox(0,0)[cc]{$c$}}
\put(14,43.75){\makebox(0,0)[cc]{$2$}}
\put(18,41){\makebox(0,0)[cc]{$c$}}
\put(18,43.75){\makebox(0,0)[cc]{$3$}}
\put(22,41){\makebox(0,0)[cc]{$b$}}
\put(22,43.75){\makebox(0,0)[cc]{$4$}}
\put(26,41){\makebox(0,0)[cc]{$b$}}
\put(26,43.75){\makebox(0,0)[cc]{$5$}}
\put(30,41){\makebox(0,0)[cc]{$d$}}
\put(30,43.75){\makebox(0,0)[cc]{$6$}}
\put(34,41){\makebox(0,0)[cc]{$d$}}
\put(34,43.75){\makebox(0,0)[cc]{$7$}}
\put(38,41){\makebox(0,0)[cc]{$a$}}
\put(38,43.75){\makebox(0,0)[cc]{$8$}}
\put(42,41){\makebox(0,0)[cc]{$f$}}
\put(46,41){\makebox(0,0)[cc]{$a$}}
\put(42,43.75){\makebox(0,0)[cc]{$9$}}
\put(46,43.75){\makebox(0,0)[cc]{$10$}}
\put(14,39.75){\line(0,-1){4.75}}
\put(18,39.75){\line(0,-1){2}}
\put(17.93,37.68){\line(-1,0){.8}}
\put(16.33,37.68){\line(-1,0){.8}}
\put(14.73,37.68){\line(-1,0){.8}}
\put(13.93,34.93){\line(-1,0){.8}}
\put(12.33,34.93){\line(-1,0){.8}}
\put(10.73,34.93){\line(-1,0){.8}}
\put(22,39.75){\line(0,-1){12}}
\put(26,39.75){\line(0,-1){5.75}}
\put(30,39.75){\line(0,-1){9}}
\put(34,39.75){\line(0,-1){3}}
\put(38,39.75){\line(0,-1){6.5}}
\put(21.93,27.68){\line(-1,0){.9412}}
\put(20.047,27.68){\line(-1,0){.9412}}
\put(18.165,27.68){\line(-1,0){.9412}}
\put(16.283,27.68){\line(-1,0){.9412}}
\put(14.4,27.68){\line(-1,0){.9412}}
\put(12.518,27.68){\line(-1,0){.9412}}
\put(10.636,27.68){\line(-1,0){.9412}}
\put(8.753,27.68){\line(-1,0){.9412}}
\put(6.871,27.68){\line(-1,0){.9412}}
\put(25.93,33.93){\line(-1,0){.8}}
\put(24.33,33.93){\line(-1,0){.8}}
\put(22.73,33.93){\line(-1,0){.8}}
\put(29.93,30.68){\line(-1,0){.8889}}
\put(28.152,30.68){\line(-1,0){.8889}}
\put(26.374,30.68){\line(-1,0){.8889}}
\put(24.596,30.68){\line(-1,0){.8889}}
\put(22.819,30.68){\line(-1,0){.8889}}
\put(33.93,36.68){\line(-1,0){.8}}
\put(32.33,36.68){\line(-1,0){.8}}
\put(30.73,36.68){\line(-1,0){.8}}
\put(37.93,33.18){\line(-1,0){.8889}}
\put(36.152,33.18){\line(-1,0){.8889}}
\put(34.374,33.18){\line(-1,0){.8889}}
\put(32.596,33.18){\line(-1,0){.8889}}
\put(30.819,33.18){\line(-1,0){.8889}}
\put(42,39.75){\line(0,-1){4.75}}
\put(41.93,34.93){\line(-1,0){.8}}
\put(40.33,34.93){\line(-1,0){.8}}
\put(38.73,34.93){\line(-1,0){.8}}
\put(22,32){\circle*{1.061}}
\put(30,35){\circle*{1.061}}
\put(10,32.75){\circle*{1.061}}
\put(6,35.5){\circle*{1.061}}
\put(6,20){\circle*{1.061}}
\put(42,38.5){\circle*{1.061}}
\put(3.75,35.5){\makebox(0,0)[cc]{$e$}}
\put(3.75,20){\makebox(0,0)[cc]{$a$}}
\put(12,32.75){\makebox(0,0)[cc]{$c$}}
\put(19.75,32){\makebox(0,0)[cc]{$b$}}
\put(32,35){\makebox(0,0)[cc]{$d$}}
\put(43.75,38.5){\makebox(0,0)[cc]{$f$}}
\multiput(21.93,27.68)(.969697,0){34}{{\rule{.4pt}{.4pt}}}
\multiput(37.93,32.93)(0,-.97222){19}{{\rule{.4pt}{.4pt}}}
\put(54,39.75){\vector(0,1){.07}}\put(54,27.75){\vector(0,-1){.07}}\multiput(53.93,27.68)(0,.92308){14}{{\rule{.4pt}{.4pt}}}
\put(6,15.5){\vector(-1,0){.07}}\put(38,15.5){\vector(1,0){.07}}\multiput(37.93,15.43)(-.969697,0){34}{{\rule{.4pt}{.4pt}}}
\put(22,17){\makebox(0,0)[cc]{$B_1^\theta$}}
\put(56,33.5){\makebox(0,0)[cc]{$H_1^\theta$}}
\end{picture}

\caption{ On this coalescent point process, the $8$-th individual is the first one whose type is the same as
  some point on lineage 0 anywhere in the past, so that {$8\in{\cal E}^\theta$ and} $B^\theta_1=8$. The maximum
  $H_1^\theta$ of the first $B_1^\theta$ branch lengths is shown. Also note that {$10\in{\cal E}^\theta$ and}
  $B^\theta_2=2$. }
\label{fig : nextbranchwithno}
\end{figure}

Set $K^\theta_0:=0$ and for $i\ge 1$, define $K^\theta_i$ as the label of the $i$-th individual in ${\cal E}^\theta$. In addition, set
$$
H^\theta_i:=\max\{H_j : K^\theta_i< j\le K^\theta_{i+1}\}
$$
and 
$$
B^\theta_i:= K^\theta_{i}-K^\theta_{i-1}.
$$
See Figure \ref{fig : nextbranchwithno} for a graphical representation of these quantities on a typical coalescent point process with mutations.

We write $(B^\theta, H^\theta)$ in lieu of $(B^\theta_1, H^\theta_1)$ and we define $W_\theta(x;\gamma)$ by
$$
W_\theta(x;\gamma):=\frac{1}{1-\EE\left(\gamma^{B^\theta}, H^\theta\le x\right)}\qquad x\ge 0, \gamma\in(0,1].
$$
We will also need the following notation
$$
W(x;\gamma):=\frac{1}{1-\gamma \PP(H\le x)}\qquad x\ge 0, \gamma\in(0,1].
$$
\begin{thm}
\label{thm : next branch with no}
The bivariate sequence $((B^\theta_i,H^\theta_i);i\ge 1)$ is a sequence of i.i.d.\ random pairs.
In addition, the following formula holds for all $x\ge 0$ and $\gamma\in(0,1]$ 
$$
W_\theta(x;\gamma)=e^{-\theta x} W(x;\gamma) +\theta \int_0^x W(y;\gamma)\,e^{-\theta y}\, dy.
$$
\end{thm}
\begin{rem}
\label{rem : w and wtheta}
Differentiating both sides of the previous equation w.r.t.\ the first variable yields
$$
dW_\theta(x;\gamma)= e^{-\theta x}\, dW(x;\gamma) .
$$
\end{rem}

\begin{rem}
The formula in the previous statement was shown in \cite{L09} in the special case $\gamma=1$.
\end{rem}
\paragraph{Proof.}

First observe that the pair $(K^\theta_1, H^\theta_1)$ does not depend on the haplotype of individual $0$, and that the $i$-th $(0,\cdot)$-type individual is also the next individual after $K^\theta_{i-1}$ with no mutation other than those carried by individual $K^\theta_{i-1}$. This ensures that $(K^\theta_i-K^\theta_{i-1},H^\theta_i)$ has the same law as $(K^\theta_1,H^\theta_1)$, and the independence between $(K^\theta_i-K^\theta_{i-1},H^\theta_i)$ and previous pairs is due to the independence of branch lengths and the fact that new mutations can only occur on branches with labels strictly greater than $K^\theta_{i-1}$.

As for the formula relating $W^\theta$ and $W$, we consider the renewal process $S$ defined by $S_0=0$ and $S_n = \sum_{i=1}^n B_i^\theta$. Next, for any integer $k\ge 0$, let $F_k$ denote the event 
$$
F_k:=\{\exists n\ge 0 : S_n = k,\ M_n \le x\},
$$
where $M_n:=\max\{H_i^\theta: 1\le i\le n\}$. Let $T_k$ denote the time elapsed since the lineages of individual 0 and individual $k$ have split up, that is, $T_k=\max\{H_i: 1\le i\le k\}$. Notice that by definition of $H_i^\theta$, $T_k=M_n$ on the event $\{S_n=k\}$, so that
$$
F_k=\{\exists n\ge 0 : S_n = k,\ T_k \le x\}.
$$
So $F_k$ is the event that the lineage of individual $k$ has had no mutation between time $-T_k$ and present time 0 (i.e.,  no mutation on the part of its lineage not common with individual 0), and $T_k\le x$. By standard properties of Poisson processes, we get
\begin{align}
\PP(F_k)&=\EE\left(e^{-\theta T_k}, T_k \le x\right) \notag \\
			&=\PP(H\le x)^k e^{-\theta x} + \theta\int_0^x   \PP(H\le y)^k\,e^{-\theta y}\, dy. \label{eq:P-F_k}
\end{align}
Note that the r.h.s.\ of this equation is obtained using the integration by parts formula for c\`adl\`ag
functions (i.e.\ functions continuous on the right and admitting left limits at each points of the space, like
$\PP(H\leq x)$\,): if $f$ is continuously differentiable and $g$ is c\`adl\`ag with
bounded variation,
\begin{equation}
  \label{eq:IPP}
f(x)g(x)=f(0)g(0)+\int_0^x f'(y)g(y)dy+\int_{(0,x]}f(y)dg(y).  
\end{equation}
Equation~(\ref{eq:P-F_k}) yields 
$$
\sum_{k\ge 0}\gamma^k\PP(F_k) =  e^{-\theta x} W(x;\gamma) + \theta\int_0^x W(y;\gamma)\,e^{-\theta y}\, dy.
$$
On the other hand,
\begin{eqnarray*}
\sum_{k\ge 0}\gamma^k\PP(F_k) &= & \sum_{k\ge 0} \gamma ^k\sum_{n\ge 0}\PP(S_n = k, M_n \le x)\\
		&=& \sum_{n\ge 0} \EE\left(\gamma^{S_n}, M_n \le x\right)\\
		&=&	\sum_{n\ge 0} \EE\left(\gamma^{\sum_{i=1}^n B_i^\theta}, H_1^\theta \le x,\ldots,H_n^\theta \le x\right)\\
		&=&	\sum_{n\ge 0} \left(\EE\left(\gamma^{B^\theta}, H^\theta\le x\right)\right)^n\\
		&=&\frac{1}{1-\EE\left(\gamma^{B^\theta}, H^\theta \le  x\right)},
\end{eqnarray*}
which yields the desired result. \hfill $\Box$

\subsection{Expected haplotype frequencies for geometrically distributed population sizes}

Let $X$ denote some independent geometric random variable with parameter $\gamma$, that is, $\PP(X\ge n)=\gamma^n$ for any $n\ge 0$.

In the infinite-allele model, each haplotype is characterized by its most recent mutation.
We denote by $A_\theta(k,y;\gamma)$ the number of haplotypes whose most recent mutation occurred between time $-y$ and present time 0 and which are carried by $k$ individuals among $\{0,1,\ldots, X\}$. 
\begin{thm}
\label{thm : expected A geom}
For all $k\ge 1$, $y> 0$, $\gamma\in(0,1]$, $u\in[0,1]$,
$$
\EE \left(u^X A_\theta (k,y;\gamma)\right) = \frac{1-\gamma}{(1-u\gamma)^2} \int_0^y dx\,\theta\,e^{-\theta x}\, \frac{1}{W_\theta(x;u\gamma)^2}\left(1-\frac{1}{W_\theta(x;u\gamma)}\right)^{k-1}.
$$
\end{thm}

Let $I_\theta'(y;\gamma)$ (resp. $I_\theta'(y;n)$) denote the number of individuals among $\{0,1,\ldots,X\}$  (resp. $\{0,1,\ldots,n\}$) whose most recent mutation appeared between time $-y$ and present time 0. 

Let $A_\theta (y;\gamma)$ (resp. $A_\theta (y;n)$) denote the number of distinct haplotypes represented in $\{0,1,\ldots,X\}$ (resp. $\{0,1,\ldots,n\}$) whose most recent mutation appeared between time $-y$ and present time 0.

Let {$\bar{G}_\theta(y;n)$} denote the \emph{probability} that two \emph{distinct} individuals randomly drawn from $\{0,1,\cdots,n\}$ \emph{share the same haplotype} and that the most recent mutation of this common haplotype appeared between time $-y$ and present time $0$.
\begin{cor}
\label{cor : IAG}
 For any integer $n\ge 1$,
$$
\EE\, I_\theta'(y;n-1) = n (1-\exp({-\theta y})),
$$
\begin{equation}
  \label{eq:for-ST}
  \EE \,A_\theta(y;n-1)  = n\int_0^y dx\,\theta\,e^{-\theta x}\, \PP(H^\theta >x)+\int_0^y dx\,\theta\,e^{-\theta x}\, \EE\left(B^\theta \wedge n, H^\theta\le x\right).
\end{equation}
and in the case where the law of $H$ has no atom
$$
\EE \,{\bar{G}_\theta}(y;n-1)  = 2\sum_{k=1}^{n-1} \frac{k(n-k)}{n(n-1)}\int_0^y \PP(H\in dx)\, \PP(H \le x)^{k-1} e^{-\theta x}\left(e^{-\theta x}-e^{-\theta y}\right).
$$
\end{cor}
\begin{rem}
The first expectation can readily be deduced from some exchangeability argument, since each individual carries
a mutation with age smaller than $y$ with probability $1-\exp(-\theta y)$ (there {is no edge effect} since the ancestral lineage is infinite).
\end{rem}
\begin{rem}
In \cite{L09}, a pathwise result was shown for the number $A_\theta(\infty,n)$ of distinct haplotypes represented in $\{0,1,\cdots, n\}$, namely
$$
\lim_{n\to\infty} n^{-1} A_\theta(\infty,n) = \int_0^\infty dx\,\theta\,e^{-\theta x}\, \PP(H^\theta >x)\qquad a.s.
$$
\end{rem}
\begin{rem}
  In the case where the law of $H$ admits atoms, the computation of $\EE\bar{G}_\theta(y;n-1)$ can be done
  following the same line as in the proof below, using the fact that $d W(x;\gamma)$ has an atomic part. The
  computation gives
  \begin{multline*}
    \EE \,{\bar{G}_\theta}(y;n-1)  = 2\sum_{k=1}^{n-1} \frac{(n-k)}{n(n-1)}\left\{ k\int_0^y \mu_H^{n.a.}(dx)\,
      \PP(H \le x)^{k-1} e^{-\theta x}\left(e^{-\theta x}-e^{-\theta y}\right)\right. \\
    \left.+\sum_{x\in[0,y]}\left(\PP(H\leq x)^{k-1}-\PP(H <x)^{k-1}\right)\,e^{-\theta x}\,\left(e^{-\theta
          x}-e^{-\theta y}\right)\right\},
  \end{multline*}
  where $\mu_H^{n.a.}$ is the non-atomic part of the law of $H$.
\end{rem}
\paragraph{Proof.} For the first expectation, taking $u=1$ in the theorem, 
$$
\EE \,I_\theta'(y;\gamma) = \EE \sum_{k\ge 1} k A_\theta (k,y;\gamma)= \frac{1-e^{-\theta y}}{1-\gamma},
$$
using repeatedly Fubini--Tonelli theorem and $\sum_{k\ge 1}k x^{k-1} = (1-x)^{-2}$ for any $x\in[0,1)$. {The
result then follows from the inversion of the generating function} using $(1-\gamma)^{-1}=\sum_{n\ge 0}(n+1) (1-\gamma)\gamma^{n}$. 

For the second expectation, 
$$
\EE \,A_\theta(y;\gamma) = \EE \sum_{k\ge 1} A_\theta (k,y;\gamma)= \frac{1}{1-\gamma} \int_0^y dx\,\theta\,e^{-\theta x}\, \frac{1}{W_\theta(x;\gamma)} = \frac{1}{1-\gamma} \int_0^y dx\,\theta\,e^{-\theta x}\, \left(1-\EE\left(\gamma^{B^\theta}, H^\theta \le x\right)\right) .
$$
Next invert the generating function as follows
\begin{align*}
\frac{1}{1-\gamma} \EE\left(\gamma^{B^\theta}, H^\theta \le x\right) 
	&=	\sum_{n\ge 0}(n+1)(1-\gamma)\gamma^n\sum_{j\ge 0} \PP(B^\theta = j, H^\theta\le x) \gamma^j\\
	&= \sum_{n\ge 0}(1-\gamma)\gamma^n \sum_{k=0}^n (n+1-k)\PP(B^\theta = k, H^\theta\le x)\\
	&= \sum_{n\ge 0}(1-\gamma)\gamma^n \EE\left(n+1-B^\theta, B^\theta \le n, H^\theta\le x\right),
\end{align*}
which entails
\begin{align*}
\EE \,A_\theta(y;n) 
	&=	 \int_0^y dx\,\theta\,e^{-\theta x}\, \left(n+1-\EE\left(n+1-B^\theta, B^\theta \le n, H^\theta\le x\right)\right)\\
	&=	 \int_0^y dx\,\theta\,e^{-\theta x}\, \left((n+1)\PP(H^\theta >x)+\EE\left(n+1 -(n+1-B^\theta)\indic{B^\theta \le n}, H^\theta\le x\right)\right)\\
	&=	 \int_0^y dx\,\theta\,e^{-\theta x}\, \left((n+1)\PP(H^\theta >x)+\EE\left((n+1)\indic{B^\theta >n} +B^\theta\indic{B^\theta \le n}, H^\theta\le x\right)\right),
\end{align*}
which yields the result.

For the third expectation, we use the fact that the {expected} number of (unordered) pairs of individuals sharing the same haplotype (younger than $y$) equals
$$
\sum_{n\ge 0} (1-\gamma) \gamma^n {\frac{n(n+1)}{2}} {\bar{G}_\theta}(y;n)={\EE[\bar{G}_\theta (y;\gamma)]},
$$
where
$$
{\bar{G}_\theta} (y;\gamma):=\sum_{k\ge 2}{\frac{k(k-1)}{2}}A_\theta (k,y;\gamma).
$$
Now since $\sum_{k\ge 2}k(k-1) x^{k-1}= 2x(1-x)^{-3}$, we get
\begin{align*}
\EE {\bar{G}_\theta} (y;\gamma) & =\frac{{1}}{1-\gamma}\int_0^y dx\,\theta\,e^{-\theta x} (W_\theta(x;\gamma) -1)\\
			& =\frac{{1}}{1-\gamma}\int_0^y dx\,\theta\,e^{-\theta x}\int_0^x e^{-\theta z} dW(z;\gamma)\\
			& =\frac{{1}}{1-\gamma}\int_0^y dW(z;\gamma)\,e^{-\theta z}\,\left( e^{-\theta {z}}-e^{-\theta y}\right)  ,
\end{align*}
where differentiation of $W$ is understood w.r.t.\ the first variable. Then we use the fact, when the law of
$H$ has no atom,
$$
dW(z;\gamma) = \frac{\gamma\PP(H\in dz)}{(1-\gamma\PP(H\le z))^2}=\PP(H\in dz)\sum_{n\ge 0} n\gamma^n \PP(H\le z)^{n-1}.
$$
The proof ends writing the product series between the last entire series and $(1-\gamma)^{-2}=\sum_{n\ge 0}(n+1)\gamma^n$.
\hfill $\Box$

Before proving the theorem, we insert a paragraph in which we state and prove a preliminary key result.

\subsubsection{A key lemma}

We denote by $\ell_{i}$ the time elapsed since the $i$-th most recent mutation on the lineage of individual 0,
also called lineage 0. Let $N_i(y;\gamma)$ denote the number of $(0,\cdot)$-type individuals in
$\{0,1\ldots,X\}$ whose \emph{most recent mutation time} in its haplotype is $\ell_{i}$ if $\ell_{i}\le y$, and $N_i(y;\gamma)=0$ otherwise.

We also define \emph{$(0,y)$-type individuals} as those individuals that have the same type as the point at time $-y$ on lineage 0. In other words, an individual is of $(0,y)$-type if the most recent mutation of its haplotype is $\ell_{i}$ for the unique $i$ such that $\ell_{i-1}\leq y<\ell_{i}$.
In the same vein, $(0,[0,y])$-type individuals are those individuals that have the same type as some point on lineage 0 at any time between time $-y$ and present time 0. 

We denote by $Z_0(y;\gamma)$ the number of $(0,y)$-type individuals of $\{0,1,\ldots,X\}$. Note that
$Z_0(y;\gamma)=N_i(y;\gamma)$ where $i$ is such that $\ell_{i-1}\leq y <\ell_i$. Also set $I_0(y;\gamma)$ the number of $(0,[0,y])$-type individuals of $\{0,1,\ldots,X\}$ and $I_0'(y;\gamma)$ the number of $(0,\cdot)$-type individuals of $\{0,1,\ldots,X\}$ whose most recent mutation appeared between time $-y$ and present time 0. Otherwise said,
$$
I_0(y;\gamma)=I_0'(y;\gamma)+Z_0(y;\gamma)\quad\mbox{ and }\quad I_0'(y;\gamma)=\sum_{i\ge 1} N_i(y;\gamma)
$$


\begin{lem}
\label{lem : key}
For all $k\ge 1$, $y> 0$, $\gamma\in(0,1]$, $u\in[0,1]$,
$$
\sum_{i\ge 1} \EE\left(u^X,N_i(y,\gamma)=k\right)=\frac{1-\gamma}{1-u\gamma} \int_0^y dz\,\theta\,e^{-\theta z}  \,\frac{W(z;u\gamma)}{W_\theta(z;u\gamma)^2}\left(1-\frac{1}{W_\theta(z;u\gamma)}\right)^{k-1}
$$
and
$$
\EE\left(u^X,Z_0(y;\gamma)=k\right)= \frac{1-\gamma}{1-u\gamma}e^{-\theta y}\,\frac{W(y;u\gamma)}{W_\theta(y;u\gamma)^2}\left(1-\frac{1}{W_\theta(y;u\gamma)}\right)^{k-1}.
$$
\end{lem}

\begin{cor}

For all $y> 0$, $\gamma\in(0,1]$, $u\in[0,1]$,
$$
\EE\left(u^X I_0'(y;\gamma)\right)=\frac{1-\gamma}{1-u\gamma} \int_0^y dz\,\theta\,e^{-\theta z}  \,W(z;u\gamma)\quad \mbox{ and }\quad 
\EE\left(u^X Z_0(y;\gamma)\right)= \frac{1-\gamma}{1-u\gamma}e^{-\theta y}\,W(y;u\gamma).
$$
\end{cor}
\paragraph{Proof.}
Use the formulae in Lemma \ref{lem : key} and Fubini--Tonelli theorem repeatedly, in particular to see that 
$$
\EE \,I_0'(y;\gamma)\,u^X=\sum_{i\ge 1}\EE\, u^X\,N_i(y;\gamma)= \sum_{i\ge 1 }  \sum_{k\ge 1}k\EE\,u^X\,\indicbis{ N_i(y;\gamma)=k}=\sum_{k\ge 1 } k \sum_{i\ge 1}\EE\,u^X\,\indicbis{ N_i(y;\gamma)=k}.
$$
The proof ends using $\sum_{k\ge 1}k x^{k-1} = (1-x)^{-2}$ for all $x\in[0,1)$. \hfill $\Box$\\
\\
Let $n$ be a non-negative integer.
In the next corollary, $Z_0(y;n)$ denotes the number of $(0,y)$-type individuals of $\{0,1,\ldots,n\}$ and $I_0'(y;n)$ the number of $(0,\cdot)$-type individuals of $\{0,1,\ldots,n\}$ whose most recent mutation appeared between time $-y$ and present time 0.
\begin{cor}
\label{cor : IZ}
For all $y> 0$ and $n\ge 0$,
$$
\EE\, I_0'(y;n)= \int_0^y dz\,\theta\,e^{-\theta z}  \,\frac{1-\PP(H\le z)^{n+1}}{\PP(H> z)}
\quad \mbox{ and }\quad 
\EE\, Z_0(y;n)= e^{-\theta y}\,\,\frac{1-\PP(H\le y)^{n+1}}{\PP(H> y)}\, .
$$
\end{cor}
\paragraph{Proof.} We use $(1-\gamma)^{-1}=\sum_{k\geq 0}\gamma^k$ along with 
$$
W(z;\gamma)=\frac{1}{1-\gamma \PP(H\le z)} = \sum_{n\ge 0} \gamma^n \PP(H\le z)^n.
$$
 Plugging these equalities into the first formula of the first corollary evaluated at $u=1$ yields

$$
\EE \, I'_0(y;\gamma)= \int_0^y dz\,\theta\,e^{-\theta z}  \frac{1}{1-\gamma}\,W(z;\gamma)=\int_0^y dz\,\theta\,e^{-\theta z}\sum_{n\geq 0}\gamma^n\sum_{k=0}^n\PP(H\leq z)^k.
$$
Inverting the generating function yields the expression proposed for $\EE\, I_0'(y;n)$. The very same line of reasoning can be applied to get $\EE\, Z_0(y;n)$.\hfill $\Box$
\begin{rem}
\label{rem : zzero}
Keeping the expression in the proof of the theorem under the shape of a sum is more informative. Indeed, differentiating each side of the equality, we then get
$$
\EE \, I'_0(dy;n)= dy\,\theta\,e^{-\theta y}  \sum_{k=0}^n\PP(H\leq y)^k,
$$
where $I_0'(dy;n)$ denotes the number of $(0,\cdot)$-type individuals of $\{0,1,\ldots,n\}$ whose most recent mutation is of age in $(y, y+dy)$. The interpretation of this new expression goes as follows. The term $\theta\, dy$ is the probability that a mutation occurred on lineage 0 in the time interval $(y, y+dy)$ backwards in time; the term $ \PP(H\leq y)^k$ is the probability that the lineage of individual $k$ split off lineage 0 more recently than $y$; the term $e^{-\theta y}$  is the probability that the lineage of individual $k$ has undergone no mutation in the last $y$ units of time.
\end{rem}
\paragraph{Proof of Lemma \ref{lem : key}.}
Set $D_1:=1$ and for $i\ge 2$,
$$
D_i:=\min\{j\ge 1 : H_j^\theta>\ell_{i-1}\}.
$$
Also recall the renewal process $S_n=\sum_{i=1}^n B^\theta_i$. 
Then we have for all $i\ge 1$
$$
N_i(y;\gamma) = \indicbis{l_i\le y}\left(\indicbis{i=1}+\sum_{j=D_{i}}^{D_{i+1} -1}\indicbis{S_j\le X} \right),
$$ 
the indicator function of $i=1$ being due to the count of individual 0 in that case.
First, we work conditionally on the values $v_i$ of the ages $\ell_i$ of mutations of lineage 0.
Using repeatedly the lack-of-memory property of $X$, we get for all $i\ge 2$ and $k\geq 1$
\begin{multline*}
\EE\left(u^X,N_i(y;\gamma)=k\mid\ell_j=v_j, j\ge 1\right)=\cdots\\
\cdots\indicbis{v_i\le y}\,\EE\left(u^{S_{D_{i}-1}}, X\ge S_{D_{i}-1}\right)\EE\left(u^{B^\theta},B^\theta \le X, H^\theta \le v_i\mid H^\theta >v_{i-1}\right)\times \cdots\\
\cdots\times\EE\left(u^{B^\theta},B^\theta \le X, H^\theta \le v_i\right)^{k-1}\left(\EE\left(u^X, B^\theta >X\right) + \EE\left(u^{X},B^\theta \le X, H^\theta > v_i\right)\right),
\end{multline*}
where the last multiplicative term equals
\begin{align*}
\EE\left(u^X, B^\theta >X\right) + \EE\left(u^{X},B^\theta \le X, H^\theta > v_i\right)
	&=\EE\left(u^X\right) - \EE\left(u^{X},B^\theta \le X, H^\theta \le v_i\right)\\
	&= 	\EE\left(u^X\right) \left(1-\EE\left(u^{B^\theta },B^\theta \le X, H^\theta \le v_i\right)\right)\\
	&=\frac{1-\gamma}{1-u\gamma}\,\left(1-\EE\left((u\gamma)^{B^\theta }, H^\theta \le v_i\right)\right)\\
	&=\frac{1-\gamma}{(1-u\gamma)W_\theta (v_i;u\gamma)} \,.
\end{align*}
Similarly for $i=1$ and $k\geq 1$,
\begin{align*}
\EE\left(u^X,N_1(y;\gamma\right)=k\mid\ell_j=v_j, j\ge 1) & = \indicbis{v_1\le y}\,\EE\left(u^{B^\theta},B^\theta
  \le X, H^\theta \le v_1\right)^{k-1}\EE\left(u^X\right)\times\cdots \\ & \qquad\qquad\qquad\qquad \cdots\times\left(1-\EE\left(u^{B^\theta
    },B^\theta \le X, H^\theta \le v_1\right)\right)\\ &
	=\indicbis{v_1\le y}\,\EE\left((u\gamma)^{B^\theta}, H^\theta \le v_1\right)^{k-1}\frac{1-\gamma}{(1-u\gamma)W_\theta (v_1;u\gamma)} \,.
\end{align*}
Now elementary probabilistic reasoning shows that for $i\ge 2$
\begin{multline*}
  \EE\left(u^{S_{D_{i}-1}},X\ge S_{D_{i}-1}\mid\ell_j=v_j, j\ge 1\right) \\
  \begin{aligned}
    & = \sum_{k\ge 1}\left(\PP(H^\theta\le v_{i-1})\right)^{k-1}\PP(H^\theta>v_{i-1})\EE\left(u^{B^\theta},B^\theta\le X\mid H^\theta \le v_{i-1}\right)^{k-1}\\
    &=	\frac{\PP(H^\theta>v_{i-1})}{1-\EE\left((u\gamma)^{B^\theta},H^\theta \le v_{i-1}\right)}
    =\PP(H^\theta>v_{i-1})W_\theta(v_{i-1};u\gamma).
  \end{aligned}
\end{multline*}
As a consequence, for all $i\ge 2$,
\begin{multline*}
\EE\left(u^X,N_i(y;\gamma)=k\mid\ell_j=v_j, j\ge 1\right)=\cdots\\
\cdots\indicbis{v_i\le y}\,\frac{1-\gamma}{1-u\gamma}\,
\frac{W_\theta(v_{i-1};u\gamma)}
{W_\theta(v_{i};u\gamma)}
\left(\EE\left((u\gamma)^{B^\theta}, H^\theta \le v_i\right)\right)^{k-1}\EE\left((u\gamma)^{B^\theta}, v_{i-1}<H^\theta \le v_{i}\right),
\end{multline*}
whereas
$$
\EE\left(u^X,N_1(y;\gamma)=k\mid\ell_j=v_j, j\ge 1\right)=\indicbis{v_1\le y}\,\frac{1-\gamma}{1-u\gamma}\,
\frac{1}
{W_\theta(v_{1};u\gamma)}\left(\EE\left((u\gamma)^{B^\theta}, H^\theta \le v_1\right)\right)^{k-1}.
$$
It is well-known that for the Poisson point process of mutations, 
$$
\PP(\ell_{i-1}\in dx, \ell_{i} \in dz)= \frac{\theta^i x^{i-2}}{(i-2)!}\,e^{-\theta z}\, dx\, dz\qquad 0<x<z, i\ge 2,
$$
so that
$$
\sum_{i\ge 2}\EE\left(u^X,N_i(y;\gamma)=k\right)=\frac{1-\gamma}{1-u\gamma}\,\sum_{i\ge 2}\int_0^y dz \int_0^z dx\,\frac{\theta^i x^{i-2}}{(i-2)!}\,e^{-\theta z}\,\frac{1}{W_\theta(z;u\gamma)}\left(1-\frac{1}{W_\theta(z;u\gamma)}\right)^{k-1}F_\theta (x,z;u\gamma),
$$
where 
\begin{equation}
\label{eqn : def Ftheta}
F_\theta (x,z;u\gamma):=W_\theta(x;u\gamma)\EE\left((u\gamma)^{B^\theta}, x<H^\theta \le z \right).
\end{equation}
Since 
$$
\EE\left(u^X,N_1(y;\gamma)=k\right)= \frac{1-\gamma}{1-u\gamma}\,\int_0^y dz \,\theta\,e^{-\theta z}\,\frac{1}{W_\theta(z;u\gamma)}\left(1-\frac{1}{W_\theta(z;u\gamma)}\right)^{k-1},
$$
we get
$$
\sum_{i\ge 1}\EE\left(u^X,N_i(y;\gamma)=k\right)=\frac{1-\gamma}{1-u\gamma}\,\int_0^y dz \,\theta\,e^{-\theta z}\,\frac{1}{W_\theta(z;u\gamma)}\left(1-\frac{1}{W_\theta(z;u\gamma)}\right)^{k-1} \left[1+\theta\int_0^z dx\,e^{\theta x}F_\theta (x,z;u\gamma)\right].
$$
Now observe that
\begin{align*}
F_\theta (x,z;u\gamma)&=W_\theta(x;u\gamma)\left(\EE\left((u\gamma)^{B^\theta}, H^\theta \le z \right)
-
\EE\left((u\gamma)^{B^\theta}, H^\theta \le x \right)\right)\\
	&= W_\theta(x;u\gamma)\left(\frac{1}{W_\theta(x;u\gamma)}-\frac{1}{W_\theta(z;u\gamma)}\right)\\
		&=1-\frac{W_\theta(x;u\gamma)}{W_\theta(z;u\gamma)}\, ,
\end{align*}
so that the integration by parts formula~(\ref{eq:IPP}) yields
$$
1+\theta\int_0^z dx\,e^{\theta x}F_\theta (x,z;u\gamma)=
1+	\left[e^{\theta x}\left(1-\frac{W_\theta(x;u\gamma)}{W_\theta(z;u\gamma)}\right)\right]_0^z +\int_0^z \frac{e^{\theta x}}{W_\theta(z;u\gamma)}\,dW_\theta(x;u\gamma),
$$
where differentiation of $W$ is understood w.r.t.\ the first variable. Since by Theorem \ref{thm : next branch with no}, $dW_\theta(x;u\gamma)=e^{-\theta x}dW(x;u\gamma)$, we get
\begin{equation}
\label{eqn : calcul Ftheta}
1+\theta\int_0^z dx\,e^{\theta x}F_\theta (x,z;u\gamma)=\frac{W(z;u\gamma)}{W_\theta(z;u\gamma)} \, ,
\end{equation}
which ends the proof for the first formula. 
Let us turn to $Z_0(y;\gamma)$. The same kind of reasoning as previously shows that
\begin{multline*}
\EE\left(u^X,Z_0(y;\gamma)=k\mid\ell_j=v_j, j\ge 1\right)=\cdots\\
\cdots\sum_{i\ge 1}\,\indicbis{v_{i-1}<y<v_i}\EE\left(u^{S_{D_{i}-1}}, X\ge S_{D_{i}-1}\right)\left(\EE\left(u^{B^\theta},B^\theta \le X, H^\theta \le y\mid H^\theta >v_{i-1}\right)\indicbis{i\ge 2}+\indicbis{i =1}\right)\times \cdots\\
\cdots\times\EE\left(u^{B^\theta},B^\theta \le X, H^\theta \le y\right)^{k-1}\left(\EE\left(u^X, B^\theta >X\right) + \EE\left(u^{X},B^\theta \le X, H^\theta > y\right)\right).
\end{multline*}
Referring to the calculations above, we easily get
\begin{multline*}
\EE\left(u^X,Z_0(y;\gamma)=k\mid\ell_j=v_j, j\ge 1\right)=\frac{1-\gamma}{1-u\gamma}\,\sum_{i\ge 1}\,\indicbis{v_{i-1}<y<v_i}\,
\frac{1}{W_\theta(y;u\gamma)}\times\cdots \\ \cdots\times\left(1-\frac{1}{W_\theta(y;u\gamma)}\right)^{k-1}
\left[\indicbis{i=1}+\indicbis{i\ge 2}W_\theta(v_{i-1};u\gamma)\EE\left((u\gamma)^{B^\theta},v_{i-1}<H^\theta\le y\right)\right].
\end{multline*}
Integrating over the law of the Poisson point process of mutations yields
\begin{multline*}
\EE\left(u^X,Z_0(y;\gamma)=k\right)= \frac{1-\gamma}{1-u\gamma}\,e^{-\theta y}
\frac{1}{W_\theta(y;u\gamma)}\left(1-\frac{1}{W_\theta(y;u\gamma)}\right)^{k-1}\\
+\frac{1-\gamma}{1-u\gamma}\,\sum_{i\ge 2}\,\int_{y}^\infty dz\int_0^y dx\,\frac{\theta^i x^{i-2}}{(i-2)!}\,e^{-\theta z}\,
\frac{1}{W_\theta(y;u\gamma)}\left(1-\frac{1}{W_\theta(y;u\gamma)}\right)^{k-1}
\,F_\theta(x,y;u\gamma),
\end{multline*}
where $F_\theta$ was defined in \eqref{eqn : def Ftheta}.
Thanks to equation \eqref{eqn : calcul Ftheta}, we get
\begin{align*}
\EE\left(u^X,Z_0(y;\gamma)=k\right)&= \frac{1-\gamma}{1-u\gamma}\,e^{-\theta y}
\frac{1}{W_\theta(y;u\gamma)}\left(1-\frac{1}{W_\theta(y;u\gamma)}\right)^{k-1}
\,\left[
1+\theta\int_0^y dx\,e^{\theta x} F(x,y;u\gamma)\right]\\
	&= \frac{1-\gamma}{1-u\gamma}\,e^{-\theta y}\,\frac{W(y;u\gamma)}{W_\theta(y;u\gamma)^2}\left(1-\frac{1}{W_\theta(y;u\gamma)}\right)^{k-1},
\end{align*}
which is the desired formula.\hfill $\Box$

\subsubsection{Proof of Theorem \ref{thm : expected A geom}}

Let $M_n(k,y;\gamma)$ denote the number of haplotypes whose most recent mutation occurred between time $-y$ and present time \emph{on the $n$-th branch} (with i.i.d.\ lengths $H_n$, except $H_0=+\infty$), and which are carried by $k$ individuals among $\{0,1,\ldots, X\}$ (hence among $\{n,n+1,\ldots, X\}$). In particular,
$$
A_\theta (k,y;\gamma) = \sum_{n\ge 0}M_n(k,y;\gamma).
$$
First,
$$
M_0(k,y;\gamma)=\sum_{i\ge 1} \indicbis{N_i(y,\gamma)=k},
$$
so thanks to Lemma \ref{lem : key}, 
$$
\EE\left(u^X M_0(k,y;\gamma)\right)= \int_0^y dz\,F(k,z;u\gamma),
$$
where we have used the following definition
$$
F(k,z;u\gamma):=\frac{1-\gamma}{1-u\gamma}\,\theta\,e^{-\theta z}  \,\frac{W(z;u\gamma)}{W_\theta(z;u\gamma)^2}\left(1-\frac{1}{W_\theta(z;u\gamma)}\right)^{k-1}.
$$
Second, for all $n\ge 1$, by the lack-of-memory property of the geometric variable $X$,
\begin{align*}
\EE\left(u^X M_n(k,y;\gamma)\right)	&= u^n \PP(X\ge n)\left[\int_0^y \PP(H_n\in dx) \EE\left(u^X M_0(k,x;\gamma)\right) + \PP(H_n\ge y)\EE\left(u^X M_0(k,y;\gamma)\right)\right]\\
	&= (u\gamma)^n \left[\int_0^y \PP(H\in dx) \int_0^x dz\,F(k,z;u\gamma) + \PP(H\ge y)\int_0^y dz\,F(k,z;u\gamma)\right]\\
		&= (u\gamma)^n \int_0^y dz\,F(k,z;u\gamma)\,\PP(H\geq z).
\end{align*}
Now since $A_\theta (k,y;\gamma) = \sum_{n\ge 0}M_n(k,y;\gamma)$, we get
\begin{align*}
\EE\left(u^X A_\theta(k,y;\gamma)\right)	&= \int_0^y dz\,F(k,z;u\gamma) +\sum_{n\ge 1}(u\gamma)^n \int_0^y dz\,F(k,z;u\gamma)\,\PP(H\geq z)\\
	&= \int_0^y dz\,F(k,z;u\gamma)\left[1+\frac{u\gamma}{1-u\gamma}\,\PP(H\geq z)\right]\\
	&= \int_0^y dz\,F(k,z;u\gamma)\left[(1-u\gamma)W(z;u\gamma)\right]^{-1},
\end{align*}
hence the result, recalling the definition of $F$.\hfill$\Box$

\section{Splitting trees: Expected haplotype frequencies at fixed time} 
\label{sec:splitting-trees}

\subsection{Joint expected haplotype frequencies with population size distribution}
\label{sec:ST-1}

In this subsection, we apply the results of the previous section to a splitting tree started at time $-t$ from one single individual and conditioned to be extant at present time 0. Then the population at present time is $\{0,1,\ldots, N_t-1\}$, where $N_t$ is the population size and $N_t-1$ follows the geometric distribution with parameter
$$
\gamma_t:=\PP(H\le t)\qquad t>0,
$$
that is, $\PP^\star(N_t-1\ge n)= \gamma_t^n$ for any integer $n\ge 0$, where $\PP^\star$ denotes the
probability conditional on the population being extant at time $0$. We recall that, in the case of splitting
trees, the law of the branch lengths $H$ is always absolutely continuous w.r.t.\ Lebesgue's measure.\\
The difference with the previous section is that the lengths of branches are (still i.i.d.\ but)
\emph{distributed as $H$ conditional on $H\le t$}. As a consequence, everything we have done in the previous
section holds for the standing population of a splitting tree founded $t$ units of time ago and conditioned
upon survival up to $t$, replacing $\gamma$ with $\gamma_t$ and $W$ with {(from Theorem \ref{thm : next branch
    with no})}
$$
W^{(t)}(x;\alpha):= \frac{1}{1-\alpha \PP(H\le x \mid H\le t)}\qquad x\in[0,t], \alpha \in(0,1].
$$
In particular we now use $W_\theta^{(t)}$ instead of $W_\theta$, with
$$
W_\theta^{(t)}(x;\alpha) = e^{-\theta x} W^{(t)}(x;\alpha)+\theta\int_0^x dy\, W^{(t)}(y;\alpha)\,e^{-\theta y} .
$$
We call a \emph{derived haplotype} a haplotype which is different from the ancestral haplotype.
Noticing that $W^{(t)} (x;u\gamma_t)=W(x;u)$, we also have $W_\theta^{(t)}(x;u\gamma_t)=W_\theta(x;u)$, where we stick to the notation from the previous section, namely,
$$
W(x;u)=\frac{1}{1-u \PP(H\le x)}\qquad x\ge 0, u\in(0,1],
$$
and (from Theorem \ref{thm : next branch with no} again)
$$
W_\theta(x;u) = e^{-\theta x} W(x;u)+\theta\int_0^x dy\, W(y;u)\,e^{-\theta y} .
$$
 Then the following statement stems readily from Theorem \ref{thm : expected A geom} and Lemma \ref{lem : key}. Recall that $W(x)= W(x;1)$ and that $W_\theta(x)= W_\theta(x;1)$.

\begin{prop}
\label{prop : exp hf}
 Let $A_\theta(k,t)$  denote the number of derived haplotypes represented by $k$ individuals in the standing population of a splitting tree founded $t$ units of time ago and  $Z_0 (t)$ the number of individuals in the standing population carrying the ancestral haplotype.
Then for all $t\ge 0$ and $u\in(0,1]$,
$$
\EE^\star \left(u^{N_t-1} A_\theta (k,t)\right) =\frac{W(t;u)^2}{W(t)} \int_0^t dx\,\theta\,e^{-\theta x}\, \frac{1}{W_\theta(x;u)^2}\left(1-\frac{1}{W_\theta(x;u)}\right)^{k-1}.
$$
and
$$
\EE^\star\left(u^{N_t-1},Z_0(t)=k\right)= \frac{W(t;u)^2}{W(t)}\,\frac{e^{-\theta t}}{W_\theta(t;u)^2}\left(1-\frac{1}{W_\theta(t;u)}\right)^{k-1}.
$$
\end{prop}
\begin{rem} Not to overload with notation, we have not considered the alleles of age less than $y$. If $A_\theta(k,y,t)$  denotes the number of derived haplotypes of age less than $y$, represented by $k$ individuals in the standing population of a splitting tree founded $t$ units of time ago, then we get the same formula as in the previous statement, but where the upper bound of the integral has changed
$$
\EE^\star \left(u^{N_t-1} A_\theta (k,y,t)\right) =\frac{W(t;u)^2}{W(t)} \int_0^{y\wedge t} dx\,\theta\,e^{-\theta x}\, \frac{1}{W_\theta(x;u)^2}\left(1-\frac{1}{W_\theta(x;u)}\right)^{k-1}.
$$
\end{rem}
The following corollary is obtained by taking $u=1$ in the last statement. A more explanatory proof is given in the next subsection.
\begin{cor}
\label{cor : expected h freq}
We have
$$
\EE^\star A_\theta (k,t) =W(t)\int_0^t dx\,\theta\,e^{-\theta x}\, \frac{1}{W_\theta(x)^2}\left(1-\frac{1}{W_\theta(x)}\right)^{k-1}
$$
and
$$
\PP^\star\left(Z_0(t)=k\right)= W(t)\,\frac{e^{-\theta t}}{W_\theta(t)^2}\left(1-\frac{1}{W_\theta(t)}\right)^{k-1}.
$$
\end{cor}
The same kinds of calculations as those done for the corollaries of the previous section yield the following statement, where the first equation could readily be deduced by exchangeability arguments.
\begin{cor}
Recall that $Z_0(t)$ is the number of individuals in the standing population carrying the ancestral type and set $A_\theta(t)$ the number of derived haplotypes represented in the standing population. Then for any positive real number $t$ and positive integer $n$,
$$
\EE (Z_0(t)\mid N_t= n )= n \exp(-\theta t)
$$
and
\begin{multline*}
\EE (A_\theta (t)\mid N_t= n ) = n\int_0^t dx\,\theta\,e^{-\theta x}\EE \left(1-\PP(H\le t)^{-B^\theta}\indic{H^\theta \le x}\right)\\
+\int_0^y dx\,\theta\,e^{-\theta x}\EE\left(\left( B^\theta\wedge n\right) \PP(H\le t)^{-B^\theta}, H^\theta\le x\right).
\end{multline*}
\end{cor}
\paragraph{Proof.} The first result is clear letting $y$ go to $+\infty$ in Corollary~\ref{cor : IZ}. In view of~(\ref{eq:for-ST}) in Corollary~\ref{cor : IAG}, in order
to prove the second result, we only need to check that
$$
\tilde{\PP}(H^\theta>x)=\EE \left(1-\PP(H\le t)^{-B^\theta}\indic{H^\theta \le x}\right)
$$
and
$$
\tilde{\EE}(B^\theta\wedge n,H^\theta\leq x)=\EE\left(\left( B^\theta\wedge n\right) \PP(H\le t)^{-B^\theta},
  H^\theta\le x\right),
$$
where $\tilde{\PP}$ is the law of the coalescent point process when the r.v.\ $(H_i)$ are i.i.d.\ with common
law $\PP(H\in\cdot\mid H\leq t)$. Now,
\begin{align*}
  \tilde{\PP}(H^\theta\leq x) & =\PP(H^\theta\leq x\mid\forall i\leq B^\theta,\ H_i\leq t) \\ & =\sum_{k\geq
    1}\PP(B^\theta=k,\ H^\theta\leq x)\PP(H\leq t)^{-k} \\ &=\EE \left(1-\PP(H\le
    t)^{-B^\theta}\indic{H^\theta \le x}\right).
\end{align*}
The second equality, very similar, is left to the reader.\hfill $\Box$\\
\\
Recall that $G_\theta(t)$ denotes the (absolute) homozygosity in the standing population, that is,
$$
G_\theta(t)= \frac{Z_0(t)(Z_0(t)-1)}{2}+\sum_{k\ge 2}\frac{k(k-1)}{2} A_\theta (k,t),
$$
then we easily get
\begin{prop}
\label{prop:homozygo}
For all $t\ge 0$ and $u\in(0,1]$,
$$
\EE^\star \left(u^{N_t-1}G_\theta(t)\right)
=\frac{W(t;u)^2}{W(t)}( W_{2\theta}(t;u)-1).
$$
\end{prop}

Note that explicit formulas can also be obtained for the expectation of the standard homozygosity $\bar{G}_\theta(t) = 2G_\theta(t)/N_t(N_t-1)$, which is the probability that two randomly sampled individuals in the population at time $t$ have the same haplotype. Formulas are given in Section~\ref{sec:Homozygosity}, where they are obtained thanks to an alternative proof based
on moment generating function computations.

\paragraph{Proof.} We use {Proposition~\ref{prop : exp hf} and the fact that $\sum_{k\ge 2}k(k-1)x^{k-2} = 2/(1-x)^{3}$.} An integration by parts yields
\begin{align*}
& \EE \left(u^{N_t-1}G_\theta(t)\right) =\frac{W(t;u)^2}{W(t)}\,e^{-\theta t}({W_\theta(t;u)}-1)+\frac{W(t;u)^2}{W(t)} \int_0^t dx\,\theta\,e^{-\theta x}\, ({W_\theta(x;u)}-1)\\
	&\quad =\frac{W(t;u)^2}{W(t)}\,e^{-\theta t}({W_\theta(t;u)}-1)+\frac{W(t;u)^2}{W(t)}\left(\Big[-e^{-\theta x}({W_\theta(x;u)}-1)\Big]_0^t+ \int_0^t dx\,e^{-\theta x}\, W_\theta'(x;u)\right),
\end{align*}
where differentiation is understood w.r.t.\ the first variable. Recalling that $W_\theta'(x;u)=e^{-\theta x}\, W'(x;u)$ provides the announced formula.\hfill $\Box$

\subsection{An explanatory proof of Corollary \ref{cor : expected h freq}}

\label{subsec : explanatory}
Consider the standing population at time $t$ conditioned on being nonempty (probability measure $\PP^\star$).
For any real number $y\in(0,t)$, for any non-negative integer $i$, let $C_i(y;dy)$, $D_i(y)$ and $E_i(y)$ denote the following events
\begin{multline*}
  C_i(y;dy):=\{i\le N_t-1,\mbox{ the $i$-th branch length has size $H_i \ge y$} \\ \mbox{ and carries a mutation with
    age in } (y, y+dy)\}  
\end{multline*}
\begin{multline*}
  D_i(y):=\{ \mbox{the type carried by the lineage of the $i$-th individual at time $t-y$} \\ \mbox{ has at least one
    alive representative} \}
\end{multline*}
$$
E_i(k,y):= \{ \mbox{the type carried by the lineage of the $i$-th individual at time $t-y$ has $k$ alive representatives} \}
$$
Then define $A_\theta (k,t,y;dy)$ as the number of haplotypes of age in the interval $(y, y+dy)$ represented by exactly $k$ alive individuals at time $t$. Hereafter, we compute the expectation under $\PP^\star$ of $A_\theta (k,t,y;dy)$. The result will follow from the equality
$$
A_\theta (k,t)=\int_0^t A_\theta (k,t,y;dy).
$$
Now it is readily seen that 
$$
 A_\theta (k,t,y;dy) =\sum_{i\ge 0}\indicbis{C_i(y;dy)\cap E_i(k,y)} 
$$
so that
$$
\EE^\star A_\theta (k,t,y;dy) =\sum_{i\ge 0} \PP^\star(C_i(y;dy)\cap E_i(k,y)). 
$$
Next observe that $E_i(k,y)\subseteq D_i(y)$, so that
\begin{align*}
\PP^\star(C_i(y;dy)\cap E_i(k,y))&= \PP^\star(C_i(y{;dy}))\PP^\star (D_i(y)\mid C_i(y;dy))\PP^\star( E_i(k,y)\mid D_i(y)\cap C_i(y;dy))\\
	&= \PP^\star(C_i(y;dy))\PP^\star (D_0(y))\PP^\star( E_0(k,y)\mid D_0(y)).
\end{align*}
Thus, we record that 
\begin{equation}
\label{eqn : 0}
\EE^\star A_\theta (k,t,y;dy) =\PP^\star (D_0(y))\PP^\star( E_0(y)\mid D_0(y))\sum_{i\ge 0} \PP^\star(C_i(y;dy)). 
\end{equation}
We will now prove the three following equalities
\begin{equation}
\label{eqn : 1}
\sum_{i\ge 0} \PP^\star(C_i(y;dy)) = \theta \,dy\ \frac{W(t)}{W(y)} ,
\end{equation}
\begin{equation}
\label{eqn : 2}
\PP^\star(D_0(y)) = \frac{W(y)\,e^{-\theta y}}{W_\theta(y)} ,  
\end{equation}
\begin{equation}
\label{eqn : 3}
\PP^\star(E_0(k,y)\mid D_0(y)) = \frac{1}{W_\theta(y)}\left(1-\frac{1}{W_\theta(y)}\right)^{k-1} .  
\end{equation}
These three equalities, along with \eqref{eqn : 0}, yield the expected expression
\begin{equation}
  \label{eq:A-with-age}
\EE^\star A_\theta (k,t,y;dy) =\theta\, dy\, W(t)\frac{e^{-\theta y}}{W_\theta(y)^2}\left(1-\frac{1}{W_\theta(y)}\right)^{k-1} ,
\end{equation}
which now sheds light on the meaning of each of the terms in the formula given in Corollary \ref{cor : expected h freq}.
Let us now prove equations \eqref{eqn : 1}, \eqref{eqn : 2} and \eqref{eqn : 3}.
First, 
\begin{align*}
\PP^\star (C_i(y;dy)) &= \PP^\star (N_t-1\ge i)\ \theta \,dy\ (\indicbis{i=0} +\indicbis{i\ge 1}\PP(H\ge y \mid H <t))\\
								&= \left(1-\frac{1}{W(t)}\right)^i \theta \,dy\ \left(\indicbis{i=0} +\indicbis{i\ge 1}\frac{\frac{1}{W(y)}-\frac{1}{W(t)}}{1-\frac{1}{W(t)}} \right)\\
								&= \theta \,dy\ \left[\indicbis{i=0} +\indicbis{i\ge 1}\left(1-\frac{1}{W(t)}\right)^{i-1} \left(\frac{1}{W(y)}-\frac{1}{W(t)}\right) \right],
\end{align*}
so we get \eqref{eqn : 1}.

Second, let $L$ denote an independent exponential r.v.\ with parameter $\theta$, so that $(y-L)^+$ is the age of the oldest mutation on lineage 0 with age smaller than $y$, with the convention that this age is zero when there is no such mutation. Then either $L\ge y$, and $D_0(y)$ is realized because lineage 0 has carried the same type since time $t-y$, or $L<y$ and $D_0(y)$ is realized iff the next branch with no extra mutation than 0 for which the maximum of past branch lengths exceeds $t-L$ satisfies that this maximum does not exceed $y$ (see Subsection \ref{subsec : next branch with no}). Conditional on $L=x$, this last event occurs with probability $\PP(H_\theta \le y\mid H_\theta >y-x)$. As a consequence, we get
\begin{eqnarray*}
\PP^\star(D_0(y)) &=& e^{-\theta y} +\int_0^y dx\,\theta\,e^{-\theta x }\left(1-\frac{W_\theta(y-x)}{W_\theta(y)}\right)\\
				&=& 1-\frac{1}{W_\theta(y)}\int_0^y dx\,\theta\,e^{-\theta x }\,W_\theta(y-x)\\
				&=& 1-\frac{e^{-\theta y }}{W_\theta(y)}\int_0^y du\,\theta\,e^{\theta u }\,W_\theta(u),
\end{eqnarray*}
and an integration by parts using the relationship between $W$ and $W_\theta$
(see Remark \ref{rem : w and wtheta}) yields \eqref{eqn : 2}.

Finally, \eqref{eqn : 3} stems from the definition of $W_\theta$ (see again Subsection \ref{subsec : next branch with no}).

\section{Splitting trees: A.s. convergence of haplotype frequencies}
\label{sec:a.s.-conv}

In this section, we rely on the theory of random characteristics introduced in the seminal papers \cite{J,N} and further developed in \cite{JNa, JNb} and especially in \cite{Taib}, where the emphasis, as here, is on branching populations experiencing mutations (but there the mutation scheme  is different, since mutation events occur simultaneously with births).

We will assume that the splitting tree starts at time $0$ with one individual. Then recall from the last
subsection that $N_t$ denotes the number of individuals alive at time $t$, $A_\theta(t)$ denotes the number of derived haplotypes carried by alive individuals at time $t$, $A_\theta(k,t)$ denotes the number of derived haplotypes carried by $k$ alive individuals at time $t$, and $Z_0(t)$ denotes the number of alive individuals at time $t$ carrying the ancestral haplotype.

For any individual $i$, in the population, we let $\chi_i(t)$ (resp.\ $\chi_i^{k}(t)$) be the number of mutations that $i$ has experienced during her lifetime that are carried by alive individuals (resp.\ by $k$ alive individuals) $t$ units of time after her birth ($\chi_i(t)=0$ if $t<0$). Then $\chi$ and the $\chi^{k}$ are individual random characteristics, in the sense given in the previously cited papers. In particular, 
$$
A_\theta(t)+Z_0(t)= \sum_i \chi_i(t-\sigma_i),
$$
and
$$
A_\theta(k,t)+\indicbis{Z_0(t)=k}= \sum_i \chi_i^{k}(t-\sigma_i),
$$
where $\sigma_i$ denotes the birth time of $i$ and the sum is taken over all individuals, dead or alive at
time $t$, in the population. This allows us to make use of limit theorems for individuals counted by random characteristics proved in \cite{J, JNa, JNb, N}, using the formulation of \cite[Appendix A]{Taib}.

Recall that $b$ is the birth rate of our homogeneous Crump--Mode--Jagers process, that $V$ denotes a random
lifetime duration, and that $\alpha$ denotes the Malthusian parameter, which satisfies $\psi(\alpha)=0$, where
$\psi$ is defined in~(\ref{eq:def-psi}). 
 
Let us restate the results in \cite[Appendix A]{Taib} in our setting.
Set 
$$
\beta:=\int_{(0,\infty]}u\, e^{-\alpha u} d\mu(u),
$$
where the last integral is a Stieltjes integral w.r.t.\ the nondecreasing function 
$$
\mu(t)=\EE(\# \mbox{ offspring born on } (0,t] )= b\EE(t\wedge V) = \int_{(0,+\infty]}(r\wedge t) \Lambda(dr) .
$$
Also for any individual random characteristic, say $\chi$, define $\widehat{\chi}(\alpha)$ as its Laplace transform at $\alpha$
$$
\widehat{\chi}(\alpha):=\int_{(0,+\infty)}dt\,e^{-\alpha t} \chi(t),
$$
where it is implicit that $\chi$ is the characteristic of the progenitor (born at time 0).
Hereafter, we apply Theorems~1 and 5 of ~\cite[Appendix A]{Taib}. These theorems need some technical assumptions to hold, which we verify at the end of the proof of the next statement. These theorems ensure first that 
$$
\lim_{t\to\infty}e^{-\alpha t} \EE A_\theta(k,t) = \frac{\EE \widehat{\chi^{k}}(\alpha)}{\beta}
$$
and second that, on the survival event,
$$
\lim_{t\to\infty} \frac{A_\theta(k,t)}{A_\theta(t)}=\frac{\EE\widehat{\chi^{k}}(\alpha)}{\EE\widehat{\chi}(\alpha)} \qquad a.s.
$$
In addition to verifying the validity of the aforementioned technical assumptions, it remains to compute the quantities $\beta$,  $\EE\widehat{\chi}(\alpha)$ and $\EE\widehat{\chi^{k}}(\alpha)$. With the following definitions,
$$
U_k:=\int_0^\infty dx\,\theta\,e^{-\theta x}\, \frac{1}{W_\theta(x)^2}\left(1-\frac{1}{W_\theta(x)}\right)^{k-1},
$$
and
$$
U:=\sum_{k\ge 1}U_k=\int_0^\infty dx\,\theta\,e^{-\theta x}\, \frac{1}{W_\theta(x)} ,
$$
we have $\beta = \psi'(\alpha)/\alpha$, $\EE\widehat{\chi^{k}}(\alpha)=U_k/b$ and of course $\EE\widehat{\chi}(\alpha) = U/b$. 
This can be recorded in the following proposition.
\begin{prop} 
\label{prop:a.s.-conv}
In the supercritical case,
\begin{equation}
  \label{eq:prop-a.s.-cv-1}
  \lim_{t\to\infty} e^{-\alpha t} \EE A_\theta(k,t) =\frac{\alpha U_k}{b\psi'(\alpha)}  
\end{equation}
and
\begin{equation}
  \label{eq:prop-a.s.-cv-2}
  \lim_{t\to\infty} e^{-\alpha t} \EE A_\theta(t) = \frac{\alpha U}{b\psi'(\alpha)}.  
\end{equation}
And on the survival event, 
$$
\lim_{t\to\infty} \frac{A_\theta(k,t)}{A_\theta(t)}=\frac{U_k}{U} \qquad a.s.
$$
\end{prop}

\begin{rem}
Note that it can be shown similarly that
$$
\lim_{t\to\infty} e^{-\alpha t} \EE N_t= \frac{\alpha}{b\psi'(\alpha)},
$$
and that, for example,
$$
\lim_{t\to\infty} \frac{A_\theta(t)}{N_t}= U \quad{\mbox{a.s.}}
$$
This is reminiscent of Theorem 3.2 in \cite{L09} where the same limit is obtained after conditioning on the population size to equal $n$ and letting $n\to\infty$. {This a.s. convergence is made possible by embedding all  populations of fixed size on the same space thanks to an infinite coalescent point process: the population of size $n$ is that generated by the first $n$ values of the coalescent point process.}
\end{rem}

\begin{rem}
\label{rem:equiv-W}
In \cite{L10}, it is proved in the supercritical case ($\alpha>0$) that the survival probability is $\alpha/b$ and that the scale function $W$ has the following asymptotic behaviour
$$
\lim_{t\to\infty} W(t) e^{-\alpha t} = \frac{1}{\psi'(\alpha)}.
$$
One could have used these two facts and the monotone convergence
theorem to recover~(\ref{eq:prop-a.s.-cv-1}) and~(\ref{eq:prop-a.s.-cv-2}) from Corollary \ref{cor : expected h freq}. In the following proof, we prefer to show the agreement with Corollary \ref{cor : expected h freq} by computing directly $\beta$,  $\EE\widehat{\chi}(\alpha)$ and $\EE\widehat{\chi^{k}}(\alpha)$.
\end{rem}

\paragraph{Proof.} Let us first prove that $\beta = \psi'(\alpha)/\alpha$.
Recalling the definition of $\beta$, we get
\begin{eqnarray*}
\beta	&=&	b\EE\int_0^\infty du\, u e^{-\alpha u} \indic{u<V}\\
			&=& \int_{(0,+\infty]} \Lambda (dr) \int_0^r du\, u e^{-\alpha u} \\
			&=&	\frac{1}{\alpha^2} \int_{(0,+\infty]} \Lambda (dr) \left(1-e^{-\alpha r}-\alpha re^{-\alpha r}\right) \\
		&=&	\frac{1}{\alpha^2} (\alpha -\psi(\alpha))-\frac{1}{\alpha}(1-\psi'(\alpha))\\
			&=&\frac{\psi'(\alpha)}{\alpha} .
\end{eqnarray*}
Next let us compute $\EE\widehat{\chi^{k}}(\alpha)$. Denote by $R^{(a,b)}_t$ the number of individuals alive at time $t$ descending clonally from the time
interval $(a,b)$. More specifically, for a progenitor individual alive on the time interval $(a,b)$ and
experiencing no mutation between times $a$ and $b$, $R^{(a,b)}_t$ is the number of individuals alive at $t$
(including possibly this progenitor) descending from those daughters of the progenitor who were born during
the time interval $(a,b)$, and that still carry the same type that the progenitor carried at time $a$. In
particular, since $W_\theta$ is the scale function associated with the clonal reproduction process 
\begin{align}
\label{eqn : Rtab}
\PP\left(R^{(a,b)}_t=k\right) &= \PP(N^\theta_{t-a}=k \mid \zeta = b-a)\nonumber\\
	&=  \PP(N^\theta_{t-a}\not=0 \mid \zeta = b-a)\PP(N^\theta_{t-a}=k \mid N^\theta_{t-a}\not=0 )\nonumber\\
	&=\left(1-\indicbis{t>b}\frac{W_\theta(t-b)}{W_\theta(t-a)}\right)\left(1-\frac{1}{W_\theta(t-a)}\right)^{k-1} \frac{1}{W_\theta(t-a)} ,
\end{align}
where $N^\theta$ is the population size process of a clonal splitting tree and $\zeta$ is the lifetime of the progenitor.
Now let us start with a progenitor with lifetime distributed as $V$ and denote by $\ell_i$ the time of the $i$-th point of a Poisson point process with intensity $\theta$ (the $i$-th mutation of the progenitor). Then
\begin{eqnarray*}
\EE \widehat{\chi^{k}}(\alpha)	&=& \EE\int_0^\infty dt\,e^{-\alpha t}\sum_{i\ge 1} \indic{\ell_i<V\wedge t}\,\mathbbm{1}\left(R^{(\ell_i,V\wedge\ell_{i+1})}_t=k\right)\\
								&=& \EE\int_0^\infty dt\,e^{-\alpha t}\sum_{i\ge 1}\int_0^\infty dz \int_0^z dy\,\frac{\theta^{i+1}y^{i-1}}{(i-1)!}\,
																\indic{y<V\wedge t}\,\mathbbm{1}\left(R^{(y,V\wedge z)}_t=k\right)\\
																&=& \EE\int_0^\infty dt\,e^{-\alpha t} \int_0^\infty dz \theta\, e^{-\theta z} \int_0^{z\wedge V\wedge t} dy\,\theta\,e^{\theta y}\,\mathbbm{1}\left(R^{(y,V\wedge z)}_t=k\right)\\
												&=& \EE\int_0^\infty dt\,e^{-\alpha t} \int_0^{ V_\theta\wedge t} dy\,\theta\,e^{\theta y}\,\mathbbm{1}\left(R^{(y,V_\theta)}_t=k\right),
\end{eqnarray*}
where $V_\theta$ denotes the minimum of $V$ and of an independent exponential r.v. with parameter $\theta$. Then
\begin{eqnarray*}
\EE \widehat{\chi^{k}}(\alpha)	
												&=& \int_0^\infty dt\,e^{-\alpha t} \int_{(0,\infty)} \PP(V_\theta\in du)
																					\int_0^{ t} dy\,\indic{y<u}\,\theta\,e^{\theta y}\,\PP\left(R^{(y,u)}_t=k\right)\\
																					&=& \int_0^\infty dt\,e^{-\alpha t} \int_{(0,\infty)} \PP(V_\theta\in du)
																					\int_0^{ t} dx\,\indic{t-x<u}\theta\,e^{\theta (t-x)}\,\PP\left(R^{(t-x,u)}_t=k\right)\\
																					&=& \int_0^\infty dx\,\theta\,e^{-\theta x}
																					\int_{(0,\infty)} \PP(V_\theta\in du) \int_x^{u+x} dt \,e^{(\theta-\alpha) t} \,\PP\left(R^{(t-x,u)}_t=k\right),
\end{eqnarray*}
which, thanks to \eqref{eqn : Rtab}, yields
\begin{eqnarray*}
\EE \widehat{\chi^{k}}(\alpha)	
												&=&  \int_0^\infty dx\,\frac{\theta\,e^{-\theta x}}{W_\theta(x)} \left(1-\frac{1}{W_\theta(x)}\right)^{k-1} 
																					\int_{(0,\infty)} \PP(V_\theta\in du) \int_x^{u+x} dt \,e^{(\theta-\alpha) t} \,\left(1-\indicbis{t>u}\frac{W_\theta(t-u)}{W_\theta(x)}\right)\\
												&=&  \int_0^\infty dx\,\frac{\theta\,e^{-\theta x}}{W_\theta(x)} \left(1-\frac{1}{W_\theta(x)}\right)^{k-1}\left({F_1(x)}-\frac{{F_2(x)}}{W_\theta(x)}\right),
\end{eqnarray*}
where
$$
{F_1(x)} :=   \int_{(0,\infty)} \PP(V_\theta\in du) \int_x^{u+x} dt \,e^{(\theta-\alpha) t}
$$
and
$$
{F_2(x)} := 	\int_{(0,\infty)} \PP(V_\theta\in du) \int_x^{u+x} dt \,e^{(\theta-\alpha) t} \indicbis{t>u}W_\theta(t-u).
$$
Let us compute {$F_1$ and $F_2$.} Set
$$
\psi_\theta(x) :=x-\int_{(0,\infty)}\left(1- e^{-rx}\right)\,b\, \PP(V_\theta\in dr)\qquad x\ge 0.
$$
Then \cite{L09} $\psi_\theta(x)=x\psi(x+\theta)/(x+\theta)$, and $1/\psi_\theta$ is the Laplace transform of $W_\theta$. Also recall that $\psi(\alpha)=0$, so that $\psi_\theta(\alpha-\theta)=0$. First, if $\theta=\alpha$, then ${F_1(x)} = \int_{(0,\infty)}u\,\PP(V_\theta\in du)= (1 - \psi_\alpha'(0+))/b=1/b$. Second, if $\theta\not=\alpha$, then
$$
{F_1(x)}=\frac{e^{(\theta-\alpha) x}}{\alpha-\theta} \int_{(0,\infty)} \PP(V_\theta\in du) \,\left(1-e^{-(\alpha-\theta)u}\right)= \frac{e^{(\theta-\alpha) x}}{b(\alpha-\theta)} (\alpha -\theta - \psi_\theta(\alpha-\theta)),
$$
so that whatever the respective values of $\alpha$ and $\theta$,
$$
{F_1(x)}=\frac{1}{b}e^{(\theta-\alpha) x}.
$$
We use Laplace transforms to compute {$F_2$.} For any $\kappa>0$,
\begin{align*}
\int_0^\infty dx\,\kappa\,e^{-\kappa x} {F_2(x)} &= \int_{(0,\infty)} \PP(V_\theta\in du) \int_u^\infty dt \,e^{(\theta-\alpha) t} W_\theta(t-u) \int_{t-u}^{t} dx\,\kappa\,e^{-\kappa x}\\
		&= \int_{(0,\infty)} \PP(V_\theta\in du) \left(e^{\kappa u}-1\right)\int_u^\infty dt \,e^{(\theta-\alpha-\kappa) t} W_\theta(t-u) \\
		&= \int_{(0,\infty)} \PP(V_\theta\in du) \left(e^{\kappa u}-1\right)e^{(\theta-\alpha-\kappa) u}\int_0^\infty ds \,e^{(\theta-\alpha-\kappa) s} W_\theta(s) \\
		&= \frac{1}{b}(\kappa+\alpha -\theta -\psi_\theta(\kappa+\alpha-\theta)-(\alpha-\theta-\psi_\theta(\alpha-\theta)))\frac{1}{\psi_\theta(\kappa+\alpha-\theta)} \\
&= \frac{\kappa}{b\psi_\theta(\kappa+\alpha-\theta)}-\frac{1}{b},
\end{align*}
so that 
$$
{F_2(x)}=\frac{1}{b}e^{(\theta-\alpha)x} W_\theta(x) - \frac{1}{b},
$$
and
$$
{F_1(x)}-\frac{{F_2(x)}}{W_\theta(x)}=\frac{1}{bW_\theta(x)} .
$$
As a consequence, we get 
$$
\EE \widehat{\chi^{k}}(\alpha)	= \int_0^\infty dx\,\frac{\theta\,e^{-\theta x}}{bW_\theta(x)^2} \left(1-\frac{1}{W_\theta(x)}\right)^{k-1},
$$
 which is the announced $U_k/b$.

 Last, let us check the technical assumptions required for Theorems~1 and 5 in \cite[Appendix A]{Taib} to hold. 
 For the first theorem, we have to check the following two requirements
 \begin{equation}
 \label{eqn : 1.1}
 \sum_{n\ge 0}
 \sup_{[n, n+1]} e^{-\alpha u} \EE\chi(u) <\infty
  \end{equation}
 \begin{equation}
 \label{eqn : 1.2}
 t\mapsto \EE \chi(t) \mbox{ is a.e. continuous.}
 \end{equation}
 For the second theorem, we have to check the following two requirements
 \begin{equation}
 \label{eqn : 2.1}
 \exists \ 0<\eta<\alpha, \ \EE\sup_{t\ge 0} e^{-\eta t} \chi(t) <\infty
  \end{equation}
 \begin{equation}
 \label{eqn : 2.2}
  \exists \ 0<\eta<\alpha, \ \hat{\mu} (\eta) <\infty. 
 \end{equation}
The following equality in distribution is easily seen
$$
\chi(t)=\sum_{i\ge 1} \indic{T_i \le t\wedge V} \indic{\sum_{j\ge 1}N_j(t-S_j)\indic{T_i <S_j< T_{i+1}\wedge t\wedge V} \in A},
$$
where $V$ is distributed as a lifetime, the $(T_i)$ are the ranked atoms of an independent Poisson point process with rate $\theta$ (mutation times), the $(S_i)$ are the ranked atoms of an independent Poisson point process with rate $b$ (birth times), the $(N_i)$ form an independent sequence of i.i.d. homogeneous, binary CMJ processes (descendances of daughters), and $A$ is taken equal to $\NN$, but can be taken equal to $\{k\}$ in the case of the random characteristic $\chi^{k}$. In any case, $\chi$ is dominated by a Poisson point process with rate $\theta$, so that $\EE \chi(t) \le \theta t$. This ensures that \eqref{eqn : 1.1} holds. As for \eqref{eqn : 1.2}, notice from the last displayed equation that $\EE\chi(t)=\sum_{i\ge 1} F_i(t)$, where
$$
F_i(t):= \int_0^t \int_{u}^{\infty}\PP(T_i \in du, T_{i+1}\in ds) \int_{[u,\infty)} \PP(V\in dr)  \PP\left(\sum_{j\ge 1}N_j(t-S_j)\indic{u<S_j<s\wedge t\wedge r} \in A\right).
$$
Because $T_i$ has a density w.r.t. Lebesgue measure, each $F_i$ is everywhere continuous on, say, $[0,t_0]$. In addition, for any $t\in[0,t_0]$, $F_i(t)\le \PP(T_i\le t)\le  \PP(T_i\le t_0)$ and  $\sum_{i\ge 1}\PP(T_i\le t_0)= \theta t_0 <\infty$, so we get continuity of $t\mapsto \EE\chi(t)$ on $[0,t_0]$ by dominated convergence. Because $t_0$ is arbitrary, $t\mapsto \EE\chi(t)$ is  continuous everywhere.
 
Let us treat the last two requirements. The last requirement \eqref{eqn : 2.2}  merely stems from the obvious inequality $\mu(t)\le bt$.
To prove \eqref{eqn : 2.1}, because $\chi$ is dominated by a Poisson point process, it suffices to show that for any Poisson point process $Y$ with rate 1, say, and for any $\eta >0$, $\EE\sup_{t\ge 0} e^{-\eta t} Y_t <\infty$.
In fact, setting $M_c(t):=e^{-\eta t} \,(Y_t+c)$, we claim that for large enough $c$, $M_c^2$ is a supermartingale.  Then using the inequality $\PP(\sup_t \, M_c^2(t)\ge z)\le c/z$, we get
$$
\PP(\sup_t \ Y_t\,e^{-\eta t}\ge y)\le \PP(\sup_t \ (Y_t+c)\,e^{-\eta t}\ge y)=\PP(\sup_t\ M_c^2(t)\ge y^2)\le \frac{c}{y^2} ,
$$
so that $\EE(\sup_t \ Y_t\, e^{-\eta t}) <\infty$. The only thing left to show is that $M_c^2$ is a supermartingale. Writing $({\cal F}_t)$ for the natural filtration of $Y$ and $P_s$ for a Poisson random variable with parameter $s$ independent of $Y_t$, we get
$$
\EE(M_c(t+s)^2 \mid {\cal F}_t) = e^{-2\eta (t+s)} \EE \left((Y_t+c + P_s)^2\right)= e^{-2\eta (t+s)}  \left((Y_t+c+s)^2 + s\right) \le M_c(t)^2,
$$
where the last inequality holds for any $s,t\ge 0$ if there is some positive $c$ (depending only on $\eta$) such that
$$
e^{-2\eta s} \left((x+s)^2+s\right) \le x^2\qquad x\ge c, s\ge 0.
$$
Then we study the function $f:s\mapsto x^2 e^{2\eta s} - (x+s)^2 - s$. Since $f''(s) = 4\eta^2x^2 e^{2\eta s} -2$, $f'$ is nondecreasing on $[0,+\infty)$ as soon as   $x^2 \ge 1/2\eta^2$. On the other hand, $f'(0)=2\eta x^2 -1 -2x$. Let $x^\star$ be the largest root of $x\mapsto 2\eta x^2 -1 -2x$. As soon as $x\ge x^\star$, $f'(0)\ge 0$. Setting $c:=\max(1/\eta\sqrt{2}, x^\star)$, as soon as $x\ge c$, $f'(0)\ge 0$ and $f'$ is nondecreasing on $[0,\infty)$, so that $f$ is nondecreasing on $[0,\infty)$. Since $f(0)=0$, we conclude that $f$ is non-negative on $[0,\infty)$, so that $M_c^2$ indeed is a supermartingale.\hfill$\Box$


\section{Expected homozygosities through moment generating functions}
\label{sec:Homozygosity}

We consider again the coalescent point process of Section~\ref{sec:coalescent}, constructed from $H_0=+\infty$
and the i.i.d.\ sequence of r.v.\ $(H_i)_{i\geq 1}$, with common law $\PP(H\in\cdot)$. Let us recall that, in
the case of splitting trees, the law of $H$ has a density w.r.t.\ Lebesgue's measure. We introduce the
derivative of $\log W(t)$:
\begin{equation}
  \label{eq:def-p_t}
  p(t)dt=\PP(H\leq t+dt\mid H>t)
  =W(t)\PP(H\in dt).
\end{equation}
For any time $t$, we consider the splitting tree obtained from $H_0,\ldots,H_{N_t-1}$, where $N_t:=\inf\{i\geq
1:H_i>t\}$. We then define the (standard) homozygosity $\bar{G}_\theta(t)$ as the probability that two
\emph{distinct} randomly sampled individuals in the population at time $t$ share the same haplotype,
and the \emph{absolute} homozygosity $G_\theta(t)$ as the number of pairs of \emph{distinct} individuals in the
population at time $t$ that share the same haplotype. Note that both of these quantities are 0 on the
event $\{N_t=1\}$, and on the complement event,
\begin{equation}
  \label{eq:rel-homozyg}
  \bar{G}_\theta(t)=\frac{2G_\theta(t)}{N_t(N_t-1)}.  
\end{equation}
The notation $G_\theta(t)$ coincides with that of Subsection~\ref{sec:ST-1}. We also recall that $Z_0(t)$ denotes the
number of individuals sharing the ancestral haplotype, defined here as the haplotype of individual 0 at time
$-t$.

Our goal in this section is to compute $\EE^\star(G_\theta(t))$ and $\EE^\star(\bar{G}_\theta(t))$ using another method than
in Section~\ref{sec:coalescent}. As in~\cite{L09}, we characterize the joint law of $(G_\theta(t),N_t,Z_0(t))$ as
time increases in a similar fashion as for branching processes, in order to obtain backward Kolmogorov
equations for moment generating functions involving these random variables. The result will then follow by
solving these equations.

\begin{prop}
  \label{prop:calc-homo-2}
  For all $t\geq 0$,
  the expected \emph{absolute} homozygosity is given by
  $$\EE^\star \left(G_\theta(t)\right)
=W(t)( W_{2\theta}(t)-1),
$$
whereas the expected \emph{standard} homozygosity is given by
  $$
  \EE^\star(\bar{G}_\theta(t))=\frac{e^{-2\theta t}(W(t)-1)}{2W(t)}+2\theta\int_0^te^{-2\theta
    s}\,\frac{W(s)-1}{W(t)-W(s)}\, \left[\frac{
\log W(t) -  \log W(s)   }{W(t)-W(s)}-\frac{1}{W(t)}\right]\,ds.
  $$
\end{prop}

\subsection{Joint dynamics of $G_\theta(t)$, $N_t$ and $Z_0(t)$}
\label{sec:joint-dyn}

Consider two splitting trees of age $t$, with respective absolute homozygosity, population size, number of
ancestral individuals and height processes $G_\theta(t)$, $N_t$, $Z_0(t)$, $(H_i)_{i\geq 0}$ and $G_\theta'(t)$, $N'_t$,
$Z'_0(t)$, $(H'_i)_{i\geq 0}$. We call \emph{merger} of these two splitting trees the splitting tree obtained
from the sequence of heights $H_0=+\infty,H_1,\ldots,H_{N_t-1},H''_0,H'_1,\ldots,H'_{N'_t-1}$, where $H''_0$
is obtained from the infinite branch $H'_0$ by cutting the part below $-t$. In addition, all the mutation
times are kept unchanged on each branch of the tree.

After this merger event, the new splitting tree has population size $N_t+N'_t$, the new number of ancestral
individuals is $Z_0(t)+Z'_0(t)$ and the new absolute homozigosity is, counting first the pairs of ancestral
individuals
\begin{align*}
  &\frac{(Z_0(t)+Z'_0(t))(Z_0(t)+Z'_0(t)-1)}{2}+G_\theta(t)-\frac{Z_0(t)(Z_0(t)-1)}{2}+G_\theta'(t)-\frac{Z'_0(t)(Z'_0(t)-1)}{2}
  \\ &=G_\theta(t)+G_\theta'(t)+Z_0(t)Z'_0(t).
\end{align*}

Now, we have $(G_\theta(0),N_0,Z_0(0))=(0,1,1)$ and, if the law of $(G_\theta(t),N_t,Z_0(t))$ is known for some $t\geq 0$,
then, on the time interval $[t,t+dt]$,
\begin{itemize}
\item either a mutation occurs on the ancestral branch, with probability $\theta\,dt$, and
  $$
  (G_\theta(t+dt),N_{t+dt},Z_0(t+dt))=(G_\theta(t),N_t,0),
  $$
\item either $H_{N_t}\in[t,t+dt]$, with probability $p(t)dt$ defined in~(\ref{eq:def-p_t}), and
  $$
  (G_\theta(t+dt),N_{t+dt},Z_0(t+dt))=(G_\theta(t)+G_\theta'(t)+Z_0(t)Z'_0(t),N_t+N'_t,Z_0(t)+Z'_0(t)),
  $$
  where $(G_\theta'(t),N'_t,Z'_0(t))$ is an i.i.d.\ copy of $(G_\theta(t),N_t,Z_0(t))$,
\item or nothing happens (the probability that two or more of the previous events occurs is $o(dt)$).
\end{itemize}
In other words, when the ancestral time $t$ increases, the process $(G_\theta(t),N_t,Z_0(t))$ jumps to $(G_\theta(t),N_t,0)$
with rate $\theta$ and to $(G_\theta(t)+G_\theta'(t)+Z_0(t)Z'_0(t),N_t+N'_t,Z_0(t)+Z'_0(t))$ with instantaneous rate $p(t)$.

Of course, the previous argument is quite informal, but it could easily be made rigorous by considering all
the possible events that could occur in the time interval $[t,t+s]$, and letting $s\rightarrow 0$. In
particular, the Kolmogorov equations of the following subsection can easily be justified this way. 

\subsection{Moment generating functions computations}
\label{sec:Laplace}

We define the moment generating functions
\begin{gather}
  L(t,u)=\EE^\star(G_\theta(t) u^{N_t-2}) \label{eq:def-V} \\
  M(t,u,v)=\EE^\star(u^{N_t-1}v^{Z_0(t)}), \label{eq:def-W}
\end{gather}
for all $u,v\in[-1,1]$ and $t\geq 0$. Since $G_\theta(t)=0$ if $N_t\leq 1$ and the quantities inside the
expectations are bounded by $N_t^2$, these functions have finite values. Our goal here is to compute explicit
expressions for these quantities.




Note that, for any i.i.d.\ triples of nonnegative r.v.\ $(G_\theta,N,Z_0)$ and $(G_\theta',N',Z_0')$,
$$
\EE((G_\theta+G_\theta'+Z_0Z'_0)u^{N+N'-2})=2\EE(G_\theta u^{N-2})\EE(u^{N})+\left(\EE(Z_0u^{N-1})\right)^2.
$$
Using this equation and the previous construction of the process, we can write the forward Kolmogorov equation
for the moment generating functions $L$ and $M$: for all $u,v\in[-1,1]$ and $t\geq 0$,
\begin{equation}
  \label{eq:EDP-1}
  \begin{cases}
    \partial_t L(t,u)=-(\theta+p(t))L(t,u)+\theta\, L(t,u)+p(t)\Big[2\,u\,L(t,u)\,M(t,u,1)+(\partial_v
    M(t,u,1))^2\Big] \\ L(0,u)=0,    
  \end{cases}
\end{equation}
and
\begin{equation}
  \label{eq:EDP-2}
  \begin{cases}
    \partial_t M(t,u,v)=-(\theta+p(t))M(t,u,v)+\theta\, M(t,u,1)+p(t)\,u\,(M(t,u,v))^2 \\
    M(0,u,v)=v.    
  \end{cases}
\end{equation}
\medskip

The explicit computation of the solutions of these equations requires several steps. First, for fixed $u$ and
$v$, the function $M(t,u,v)$ is solution to an ODE known as Riccati's equation. In the case where $v=1$, the
function $f(t)=M(t,u,1)$ is solution to
$$
\dot{f}=pf(uf-1),
$$
which is known as Bernoulli's equation. It can be solved by making the change of unknown function
$\tilde{f}=1/f$, which makes the ODE linear. This yields
\begin{equation}
  \label{eq:f-MGF}
  f(t)=M(t,u,1)=\left(u+(1-u)\exp\int_0^tp(s)ds\right)^{-1}=\frac{W(t;u)}{W(t)},  
\end{equation}
where we used that $p$ is the derivative of the function $\log W(t)$.

Second, for all $u,v\in[-1,1]$, the function $M(t,u,1)$ is a particular solution of~(\ref{eq:EDP-2}) (with
different initial condition). Hence, the function $g(t)=M(t,u,v)-M(t,u,1)=M(t,u,v)-f(t)$ solves the Bernoulli
ODE
$$
\dot{g}=-(\theta+p-2upf)g+upg^2,
$$
for which the previous trick again works. This yields
\begin{equation*}
  M(t,u,v)=f(t)+\frac{\exp\left(-\int_0^t(\theta+p(s)-2up(s)f(s))ds\right)}
  {(v-1)^{-1}-u\int_0^tp(s)\exp\Big(-\int_0^s(\theta+p(\tau)-2up(\tau)f(\tau))d\tau\Big)ds}.
\end{equation*}
Since $uW(s;u)\PP(H\in ds)$ is the derivative of $\log W(\cdot;u)$, it follows from~(\ref{eq:f-MGF}) that
\begin{equation}
  \label{eq:calcul} 
  \int_0^tp(s)(1-2uf(s))ds=\log W(t)-2\log W(t;u).
\end{equation}
Hence, we obtain
$$
M(t,u,v)=\frac{W(t;u)}{W(t)}\left(1+\frac{e^{-\theta
      t}\,W(t;u)}{(v-1)^{-1}-u\int_0^te^{-\theta s}W(s;u)^2\PP(H\in ds)}\right).
$$
Observing that $uW(s;u)^2\PP(H\in ds)$ is the derivative of $W(\cdot;u)$, an integration by parts and
Theorem~\ref{thm : next branch with no} finally yield
$$
M(t,u,v)=\frac{W(t;u)}{W(t)}\left(1-\frac{e^{-\theta t}\,W(t;u)}
{\frac{v}{1-v}+W_\theta(t;u)}\right).
$$
We then compute
$$
M(t,u,1)=\frac{W(t;u)}{W(t)}=f(t)\quad\hbox{and}\quad
\partial_v M(t,u,1)=\frac{W(t;u)^2\,e^{-\theta t}}{W(t)}=:q(t).
$$

Third, the linear equation~(\ref{eq:EDP-1}) can be explicitly solved:
$$
L(t,u)=\exp\left(-\int_0^tp(s)(1-2uf(s))ds\right)\,
\int_0^tp(s)q^2(s)\exp\left(\int_0^sp(\tau)(1-2uf(\tau))d\tau\right)ds.
$$
Using~(\ref{eq:calcul}) again, we obtain
$$
L(t,u)=\frac{W(t;u)^2}{W(t)}\int_0^t e^{-2\theta s}\,W(s;u)^2\,\PP(H\in ds).
$$
Using integration by parts as above finally yields
\begin{equation}
  \label{eq:finale}
  L(t,u)=\frac{W(t;u)^2}{W(t)}\,\frac{W_{2\theta}(t;u)-1}{u},
\end{equation}
which is consistent with Proposition~\ref{prop:homozygo}.



Fourth, using Theorem~\ref{thm : next branch with no}, we have
$$
\frac{W_{2\theta}(t;u)-1}{u}=e^{-2\theta t}\,\frac{W(t;u)-1}{u}+2\theta\int_0^t e^{-2\theta
  s}\,\frac{W(s;u)-1}{u}du.
$$
This yields
$$
L(t,u)=\frac{W(t;u)^2}{W(t)}\left[e^{-2\theta t}\,\PP(H\leq t)\,W(t;u)+2\theta\int_0^t e^{-2\theta
    s}\,\PP(H\leq s)\,W(s;u)ds\right].
$$

Writing the product series of $(1-v)^{-1}=\sum_{n\geq 0}v^n$ and $(1-v)^{-2}=\sum_{n\geq 0}(n+1)v^n$ and
observing that
$$
\sum_{k=0}^n(k+1)a^kb^{n-k}=\frac{d}{da}\left(a\sum_{k=0}^na^kb^{n-k}\right)
=\frac{(n+1)a^{n+2}-(n+2)a^{n+1}b+b^{n+2}}{(a-b)^2},
$$
we get
\begin{multline}
  L(t,u)=\frac{e^{-2\theta t}\PP(H\leq t)}{2 W(t)}\sum_{n\geq 2}n(n-1)(\PP(H\leq t)u)^{n-2}
  +\frac{2\theta}{W(t)}\int_0^tds e^{-2\theta s}\PP(H\leq s)\times \\ \sum_{n\geq 0 u^2}\frac{(n+1)\PP(H\leq
    t)^{n+2}-(n+2)\PP(H\leq t)^{n+1}\PP(H\leq s)+\PP(H\leq s)^{n+2}}{\PP(s< H\leq t)^2}\,u^n. \label{eq:last-eq}
\end{multline}

Finally, we compute the expected standard homozygosity as follows: by~(\ref{eq:rel-homozyg}),
$$
\partial^2_u\left(\EE\left(\bar{G}_\theta(t)u^{N_t}\right)\right)=L(t,u),\quad\text{or}\quad
\EE(\bar{G}_\theta(t))=\int_0^1 du\int_0^u dv\,L(t,v).
$$
Integrating~(\ref{eq:last-eq}) twice and using the equation
$$
(1-x)\log(1-x)+x=\sum_{n\geq 2}\frac{x^n}{n(n-1)}
$$
yields
$$
\EE^\star[\bar{G}_\theta(t)]=\frac{e^{-2\theta t}(W(t)-1)}{2W(t)}+2\theta\int_0^tds\,e^{-2\theta
  s}\,\frac{W(s)-1}{W(t)-W(s)}\,
\left[\frac{\log\frac{W(t)}{W(s)}}{W(t)-W(s)}-\frac{1}{W(t)}\right],
$$
which ends the proof of Proposition~\ref{prop:calc-homo-2}.

\paragraph{Acknowledgments.} This work was partly funded by project BLAN06-$3\textunderscore 146282$ MAEV `Modèles Aléatoires de l'\'Evolution du Vivant'  of ANR (French national research agency).

\end{document}